\documentclass[aps,prl,reprint,superscriptaddress]{revtex4-1}

\usepackage{amsmath,amsfonts,amssymb,mathtools,braket,graphicx}
\usepackage[utf8]{inputenc}
\usepackage{layouts}
\usepackage[colorlinks=true,citecolor=blue,linkcolor=red,urlcolor=blue]{hyperref}
\usepackage[normalem]{ulem}
\begin{document}
  
  \title{Quantum interference tuning of spin-orbit coupling in twisted van der Waals trilayers}

  \author{Csaba G.~P\'eterfalvi}
  \email{csaba.peterfalvi@uni-konstanz.de}
  \affiliation{Department of Physics, University of Konstanz, D-78464 Konstanz, Germany}
  
  \author{Alessandro David}
  \affiliation{Peter Gr\"unberg Institute – Quantum Control (PGI-8), Forschungszentrum J\"ulich GmbH, J\"ulich, Germany}
  
  \author{P\'eter Rakyta}
  \affiliation{Department of Physics of Complex Systems, E\"otv\"os Lor\'and University, Budapest, Hungary}
  \affiliation{Quantum Information National Laboratory, Hungary}
  
  \author{Guido Burkard}
  \email{guido.burkard@uni-konstanz.de}
  \affiliation{Department of Physics, University of Konstanz, D-78464 Konstanz, Germany}
  
  \author{Andor Korm\'anyos}
  \email{andor.kormanyos@ttk.elte.hu}
  \affiliation{Department of Physics of Complex Systems, E\"otv\"os Lor\'and University, Budapest, Hungary}

  \begin{abstract}
  We show that in  van der Waals stacks  of twisted hexagonal layers  the proximity induced  Rashba spin-orbit coupling can be affected by quantum interference. We calculate the quantum phase responsible for this effect in graphene--transition metal dichalcogenide bilayers as a function of interlayer twist angle.  We show how this quantum phase affects the spin-polarization of the graphene bands and discuss its potential effect on spin-to-charge conversion measurements.  
  In twisted trilayers symmetries can be broken as well as restored for certain twist angles. This can be used to deduce  the effects of 
    induced spin-orbit coupling on spin-lifetime anisotropy and magnetoconductance  measurements.
  \end{abstract}

  \maketitle

Multilayer stacks  of two-dimensional (2D) materials, commonly referred to as van der Waals (vdW) heterostructures, have become an important platform for exploring a wide range
of exciting phenomena, because they offer electronic tunability and  a large parameter space.
Initially, bilayer structures served as a platform to study e.g., the effects of a moire potential in graphene/hBN
\cite{Yankowitz2012,Ponomarenko2013,Hunt-Science1427,ribeiro-palau_twistable_2018}
the induced  spin-orbit coupling (SOC) in graphene/transition metal dichalcogenide(TMDC) heterostructures\cite{wang_strong-SOC_2015,wang_origin-of-SOC_2016,yang_tunable-SOC_2016}, 
or strong electronic correlations in magic-angle twisted bilayer graphene (MATG)\cite{bistritzer_magic-angle_2011,cao_unconventional_2018}. 
A natural next step is to add another layer, which introduces a second twist angle and/or a further layer of different properties and this 
can  enhance the parameter space to tune the properties of the system in several ways. 
An exciting opportunity is to use MATG and induce SOC into the flat-bands by adding a TMDC layer to the stack\cite{Arora2020,lin2021-proximity-MATG}. %It has been shown that the superconductivity in MTG is stabilized by the proximity induced SOC of a WSe2 layer\cite{}.
Further examples include  the recent proposal of an engineered topological phase in WSe$_2$/bilayer graphene (BLG)/WSe$_2$ system\cite{Island2019}, the external gate tunability of SOC through changing layer polarization  in TMDC/BLG structures\cite{BLG-proximitySOC-Levitov,BLG-proximitySOC-Gmitra,BLG-proximitySOC-Bockrath,BLG-spinvalve-Aveek} and 
imprinting double moire potential on graphene\cite{double-moire1,double-moire2,Makk-double-moire}.
More generally, vdW multilayers can serve as quantum simulator platform for strongly correlated physics and topological materials\cite{Kennes-quantum-simulator2021}.

 The proximity induced SOC plays a pivotal role in many of the  above proposals. A number of recent experimental works  
 based on weak antilocalization (WAL) \cite{wang_strong-SOC_2015,wang_origin-of-SOC_2016,yang_tunable-SOC_2016,yang_tunable-SOC_2016,yang_strong-SOC_2017,volkl_magnetotransport_2017,zihlmann_large-SOC_2018, wakamura_strong_2018,wakamura_PRB2019}
 spin-lifetime anisotropy measurements\cite{ghiasi_large_2017,vanWees_spinanisotropy-PRL-2018,benitez_strong-anistropy-2018,xu_BLG_spinanistropy_2018}, 
 spin-Hall\cite{safeer_room-temperature_2019,spin-Hall-Casanova,Benitez2020,Dash-TaTe2} and Rashba-Edelstein effect\cite{Ghiasi2019,Khokhriakov2020,Benitez2020,Ferreira_ACSNano2020} proved that SOC is strongly enhanced in graphene/TMDC heterostructures, which also motivated theoretical work to understand these measurements\cite{wang_strong-SOC_2015,Gmitra_TMDC-proximity_2016,raimondi-spin_Hall_graphene,cummings_anisotropy_2017,roche_spin-Hall-and-localization2017,ferreira-charge_to_spin2017,garcia_spin_transport_2018,ferreira-charge_to_spin2017,Gani_SOC-nanoribbon,Ferreira-electrical-detection2020,Zollner2021bilayer}. 
 Recently, Refs.~\cite{alsharari_topological_2018,li_twist-angle_2019,alessandro-SOC-and-twist,Pezo2020twist} have also discussed the interlayer  twist angle dependence of the induced SOC in graphene/TMDC.  In particular, it has been noted\cite{li_twist-angle_2019,alessandro-SOC-and-twist} that the most general form of the induced Rashba-SOC in twisted graphene/TMDC heterostructures that obeys time reversal $\mathcal{T}$ and three-fold rotation $C_3$  symmetries can be written as 
  \begin{equation}
   H_{R}^{}=\frac{\lambda^{}_R}{2} e^{i \frac{s_z}{2} {\vartheta_{R}}}\left(\tau_z\sigma_x s_y - \sigma_y s_x\right) 
  e^{-i \frac{s_z}{2} {\vartheta_{R}^{}}},
  \label{eq:HgrR-general}
  \end{equation}
where $s_{x,y,z}$ ($\sigma_{x,y,z}$) are Pauli matrices acting in the spin (sublattice) space, and $\tau_z$ is a Pauli matrix acting on the valley degree of freedom  of graphene. 
Both $\vartheta_{R}$ and  $\lambda^{}_R$ are  functions of the interlayer twist angle $\theta$, which we do not show explicitly in order to ease the notations.  
$H_{R}$ differs from the usual 
Rashba SOC term\cite{kane_quantum-spin-Hall_2005,min_intrinsic_2006,TB-SOC-graphene} $H_{R}^{}=\frac{\lambda^{}_R}{2} \left(\tau_z\sigma_x s_y - \sigma_y s_x\right) $
by a rotation of angle $\vartheta_{R}$  in  spin-space. 
The terms containing $\vartheta_R$ appear  because 
for a general interlayer twist angle the symmetry of the heterostructure is lowered from $C_{3v}$ to $C_{3}$.  
Thus  Eq.~(\ref{eq:HgrR-general}) is valid not only for graphene/TMDC heterostructures, but for a wide range of twisted  heterostructures consisting of  hexagonal layers, 
such as heterostructures of graphene with  
semiconductor\cite{Gr-on-hexagonal,Gr-on-InSe}, ferromagnetic\cite{Zollner-spin-orbit-torque}, and topological insulator\cite{Zollner_Gr-topological2019,nemes-incze_jacutingaite}  
structures. 
%Moreover, a similar phase factor appears  when a TMDC monolayer is placed in a perpendicular electric field and would also be present e.g., in  twisted heterobilayers of TMDCs.  
This general nature of the effect is one of the main motivations for our work. 
However, the physical significance of this spin-space rotation was not previously appreciated and the relation of  
$\vartheta_{R}$ to the interlayer twist angle $\theta$ has  not been discussed. 
As we will show, using graphene/TMDC twisted bilayers as an example,   $\vartheta_{R}$ can take finite values and it leads to quantum interference affecting the induced Rashba type SOC  in twisted trilayers. While it has been known that $\lambda_R$ is tunable e.g., 
by pressure\cite{Fulop2021boosting} in proximity structures, to our knowledge  the possibility that quantum interference can affect its value has not been considered before.

 \begin{figure*}[ht!]
  \includegraphics[scale=0.2]{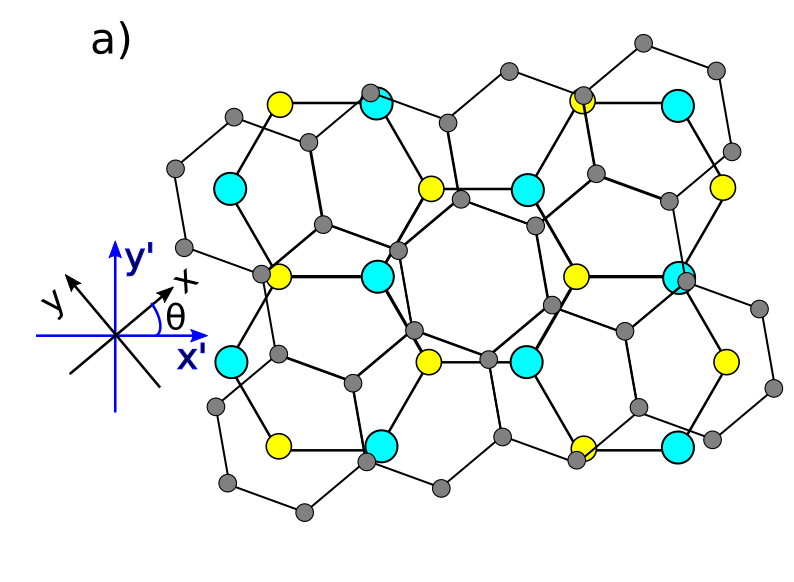}
 \includegraphics[scale=0.25]{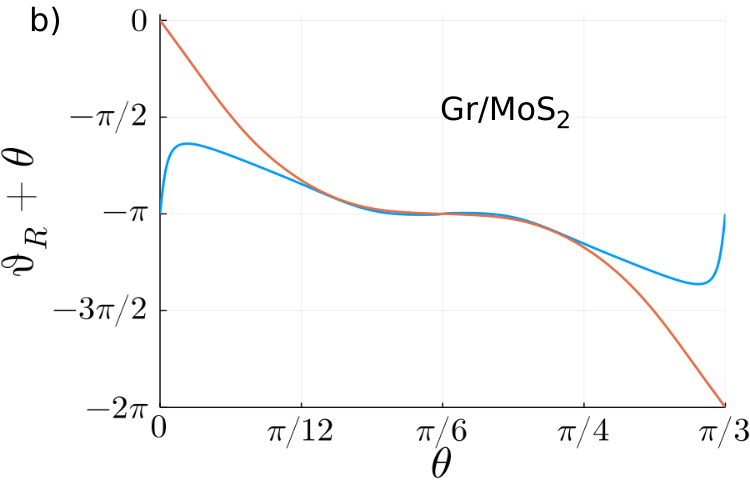}
 \includegraphics[scale=0.25]{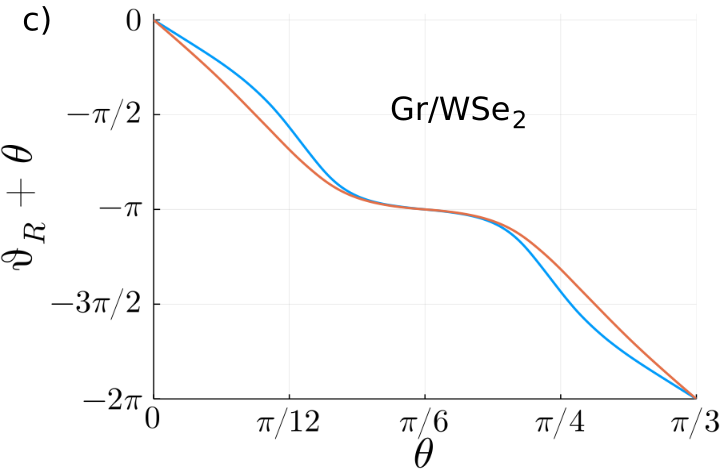}\\
\includegraphics[scale=0.3]{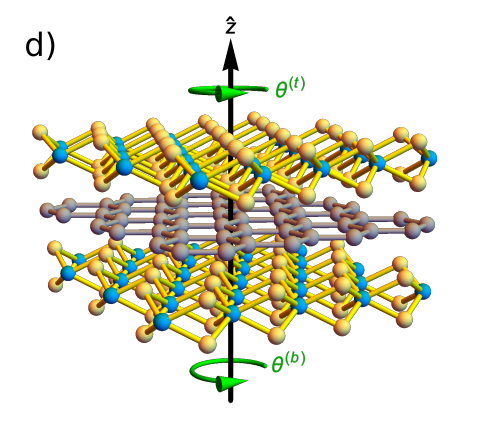}
 \includegraphics[scale=0.45]{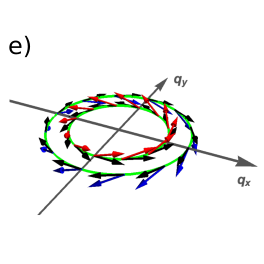}
 \includegraphics[scale=0.45]{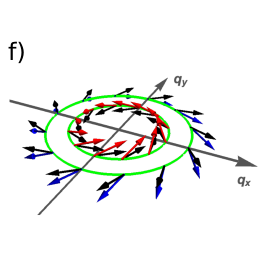}
 \includegraphics[scale=0.45]{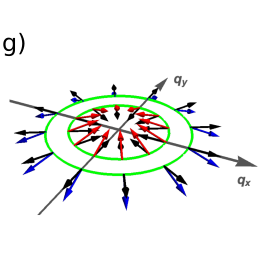}
 \caption{a) Schematics of a twisted graphene/TMDC bilayer with twist angle $\theta$. b) and c) $\vartheta_{R}+\theta$ as a function of $\theta$ for MoS$_2$ and WSe$_2$. 
           Blue (red) curves were calculated using  DFT (ARPES measurements) parameters, see main text for details. 
           d) Schematics of a twisted TMDC/graphene/TMDC trilayer.
          The spin-orbit field $\mathbf{s}(\mathbf{q})$ 
          for e) $\theta=0$,  f) $\theta=7^{\circ}$ and g) for $\theta=15^{\circ}$, using the $\vartheta_{R}$ vs $\theta$ dependence  shown in Fig.~\ref{fig:rashba-phase}(c), calculated for graphene/WSe$_2$.
          The green circles indicate the Fermi surfaces for the two spin subbands.
$\mathbf{s}(\mathbf{q})$ lies in the plane if only Rashba SOC is induced in graphene (black arrows), whereas  it  acquires a non-zero $\langle\hat{s}_z\rangle$ component if the  
valley Zeeman type SOC  is finite as well (blue and red). 
 \label{fig:rashba-phase}}
\end{figure*}

\textit{Twisted bilayers} 
In order to  obtain the angle $\vartheta_{R}$ we use the methodology 
recently developed to calculate the coupling strength $\lambda_R$ for graphene/TMDC bilayers\cite{alessandro-SOC-and-twist}.  
$\vartheta_{R}$ is a quantum phase which depends on  the
interlayer tunneling between the Bloch states of graphene and the TMDC layer and  on certain off-diagonal matrix elements of the intrinsic SOC of the TMDC, 
for details we refer to Ref.~\cite{supplementary_info}.
One finds that $\lambda_{R}$ and $\vartheta_{R}$ are the absolute value and the phase of the complex Rashba coefficient 
$\bar{\lambda}_R=\lambda_{R}e^{i \vartheta_\text{R}}$, which is  given by the sum of the contributions from pairs of even ($e$) and odd ($o$) bands of the TMDC: 
%$\lambda_R=|\bar{\lambda}_R|$ and %$\vartheta_\text{R}=\mathrm{Arg}(\bar{\lambda}_R),$ where
$\bar{\lambda}_R= \sum_q \lambda_{\text{R},(e,o)_q} e^{i\vartheta_{(e,o)_q}} = 
%\sum_{q \in \{(e,o)\}} \lambda_{\text{R},q} e^{i\vartheta_q}
\sum_{q} \lambda_{\text{R},q} e^{i\vartheta_q}.
$ 
Here $\lambda_{\text{R},q}$ and $\vartheta_q$ are the magnitude and the phase of the contributions of the pairs of bands.
In the calculations of $\lambda_{\text{R},q}$ and $\vartheta_q$
we have used the tight-binding model of Ref.~\cite{fang_ab_2015}.
For the initial steps of the calculations it is convenient to assume that the graphene layer is rotated with respect to the TMDC layer\cite{alessandro-SOC-and-twist}. 
In the final steps, we change the representation of the Hamiltonian from the coordinate system fixed to the TMDC layer to the system fixed to the graphene layer with a transformation 
$e^{-i\tau_z \frac{\sigma_z}{2}\theta} H_{R}^{} e^{i\tau_z \frac{\sigma_z}{2}\theta}$,  see Fig.~\ref{fig:rashba-phase}(a).
From the explicit form of the transformed Rashba Hamiltonian one finds that the non-zero matrix elements are  
 $\propto \lambda_R e^{\pm i(\vartheta_{R}+\theta)}$,  i.e.,  the sum of the geometric angle $\theta$ and the quantum phase $\vartheta_{R}$ 
plays an important role. 

Regarding our numerical calculations, the first ingredient is the tight-binding (TB) Hamiltonian of  Ref.~\cite{fang_ab_2015} for TMDCs. This TB model itself is derived from density functional theory (DFT) calculations and we use it,  
among others, to calculate matrix elements of the spin-orbit coupling Hamiltonian and the interlayer tunneling amplitude.  The results also depend  on i) the position of the Dirac point of graphene within the band gap $E_g$ of the TMDC, and 
ii) on the value of $E_g$.  Numerical  DFT calculations are known to often underestimate $E_g$ and they seem to give\cite{Gmitra_TMDC-proximity_2016} different results 
from experiments\cite{pierucci_graphene-MoS2,nakamura_graphene-WSe2} for  the  energy alignment of graphene's Dirac point with the TMDC bands.
We performed our calculations for two parameter sets to assess  how sensitive are the results on  the choice of these material parameters.
Since Ref.~\cite{fang_ab_2015} does not provide information on the band alignment of graphene and the TMDC layer,  we use the DFT calculations of Ref.~\cite{Gmitra_TMDC-proximity_2016} for this purpose.  The second parameter set for $E_g$ and the band alignment is extracted from ARPES 
measurements\cite{pierucci_graphene-MoS2,nakamura_graphene-WSe2}.
Since the coupling between the layers is weak at the Dirac point of graphene, the band alignment should mainly depend on the work function 
difference between the two materials. Therefore we assume that it does not depend on the interlayer twist angle. 
This is in agreement with the recent computational work of Ref.~\onlinecite{Naimer2021twistangle}.

The results  for the  $\vartheta_{R}+\theta$ vs $\theta$  dependence for two selected TMDCs, MoS$_2$ and WSe$_2$, 
are shown in Figs.~\ref{fig:rashba-phase}(b) and (c), respectively. 
In the case of  MoS$_2$, using the DFT parameter set,  one can see that $\vartheta_{R}+\theta$ remains in a limited range around $\pi$
as $\theta$ varies from $0$ to $\pi/3$ (Fig.~\ref{fig:rashba-phase}(b)). However, if parameters extracted from ARPES measurements\cite{pierucci_graphene-MoS2} 
are used then  $\vartheta_{R}+\theta$ covers the entire range $[0,2\pi]$.
For  WSe$_2$ one finds that $\vartheta_{R}+\theta$ covers all of  $[0,2\pi]$ (Fig.~\ref{fig:rashba-phase}(c)) 
and the results obtained from the two parameter sets qualitatively agree. 
The difference between the results for the two materials can be mainly traced back to the different energy alignment  of the Dirac point in the TMDC band gap.  
One can also note  in Figs.~\ref{fig:rashba-phase}(b,c) that  for $\theta=l\pi/6$, $l=0,1,\dots$ one finds $\vartheta_{R}+\theta=n\pi$, where $n$ is an integer. 
We give a detailed discussion of how this result  for $\vartheta_{R}+\theta$ follows from our theoretical method in Ref.~\cite{supplementary_info}, 
but already note at this point that for $\theta=l\pi/6$ the  vertical mirror planes of the graphene and the TMDC lattice line up  and the system, as a whole, has $C_{3v}$ symmetry. 
In this case  the Hamiltonian of the  induced Rashba SOC reads $H_{R}^{}=(-1)^{n+1}\frac{\lambda^{}_R(\theta)}{2} \left(\tau_z\sigma_x s_y - \sigma_y s_x\right)$, 
hence    $H_{R}^{}$ simplifies to the form used previously in the literature\cite{kane_quantum-spin-Hall_2005,Gmitra_TMDC-proximity_2016}. 
We find that  $n$ can be, in general, both odd  and even (see Figs.~\ref{fig:rashba-phase}(b),(c)),  which means  that $\lambda^{}_R(\theta)$ can  acquire a negative sign  as $\theta$ is changed. 
Interestingly, when $\vartheta_{R}+\theta\neq n\pi$, the spin-orbit field 
$\mathbf{s}(\mathbf{q})=(\langle\hat{s}_x\rangle,\langle\hat{s}_y\rangle,\langle\hat{s}_z\rangle)^{T}$ is not tangential to the Fermi surface 
as in the case of usual Rashba SOC (cf. Fig.~\ref{fig:rashba-phase}(e) and Figs.~\ref{fig:rashba-phase}(f),(g)). Instead, one can show that the in-plane component 
$ (\langle\hat{s}_x\rangle,  \langle\hat{s}_y\rangle)^{T}$ is rotated by an angle $\vartheta_{R}+\theta$ with respect to the tangential direction. As we will discuss, this might have consequences on the interpretation of  experimental results.

\textit{Twisted TMDC/graphene/TMDC trilayers} 
Adding another TMDC layer, as shown in Fig.~\ref{fig:rashba-phase}(d), introduces a second interlayer twist angle  
and the two  twist angles $\theta^{(b)}$ and $\theta^{(t)}$ for the bottom and top TMDC layers allow an even broader  
control of the induced SOC in graphene.

Since the layers are only weakly coupled, the effective graphene Hamiltonian  is 
$H_{orb}+H_{soc}^{(t)}+H_{soc}^{(b)}$, where $H_{soc}^{(b)}=H_{vZ}^{(b)}+H_{R}^{(b)}$ and $H_{soc}^{(t)}=H_{vZ}^{(t)}-H_{R}^{(t)}$.
Here $H_{vZ}^{}=\lambda_{vZ}^{}\tau_z s_z$ is the Hamiltonian of the induced 
valley Zeeman SOC in the graphene layer.
Note, that the contributions $H_{R}^{(b)}$ and $H_{R}^{(t)}$ have a different sign. 
As a simple physical explanation, consider the case when the two TMDC layers are perfectly aligned, e.g., $\theta^{(b)}=\theta^{(t)} $. 
Then the graphene layer is  horizontal mirror plane of the whole stack,  which dictates  that the Rashba SOC must vanish.  
(A more microscopic argument is given in Ref.~\cite{supplementary_info}.) 
One can define the complex Rashba coefficient for the trilayer system (tls) by
\begin{eqnarray}
 \bar{\lambda}_R^{(tls)}
 %&=&\bar{\lambda}_R^{(b)}e^{i\theta^{(b)}}-\bar{\lambda}_R^{(t)}e^{i\theta^{(t)}} \nonumber \\
 =\lambda_R^{(b)}e^{i\left(\vartheta_{R}^{(b)}+\theta^{(b)}\right)}-\lambda_R^{(t)}e^{i\left(\vartheta_{R}^{(t)}+\theta^{(t)}\right)},
 \label{eq:trilayer-complex-lambda_R}
\end{eqnarray}
and its magnitude ${\lambda}_R^{(tls)}=\left|\bar{\lambda}_R^{(tls)}\right|$ and phase $\vartheta^{(tls)}={\rm Arg}\left[\bar{\lambda}_R^{(tls)}\right]$.
In terms of these quantities the  induced Rashba type SOC can be written as  
$H_{R}^{(tls)}%=H_{R}^{(b)}-H_{R}^{(t)}
=\frac{\lambda^{(tls)}_R}{2} 
 e^{i \frac{s_z}{2} {\vartheta^{(tls)}}}
\left(\tau_z\sigma_x s_y - \sigma_y s_x\right)
e^{-i \frac{s_z}{2} {\vartheta^{(tls)}}}
$.
The importance of the phase $e^{i(\vartheta_{R}+\theta)}$ discussed for bilayers becomes now more  clear: 
it follows from Eq.~(\ref{eq:trilayer-complex-lambda_R}) that  the strength ${\lambda}_R^{(tls)}$ of the induced Rashba SOC in trilayer stacks can be affected 
by quantum interference effects if $\vartheta_{R}^{(b)}+\theta^{(b)}$ and/or $\vartheta_{R}^{(t)}+\theta^{(t)}$ are non-zero.
This can be interpreted as an interference of the  virtual hopping processes to the two TMDC layers 
that give rise to the induced  Rashba SOC.
\begin{figure}[t!]
  \includegraphics[scale=0.20]{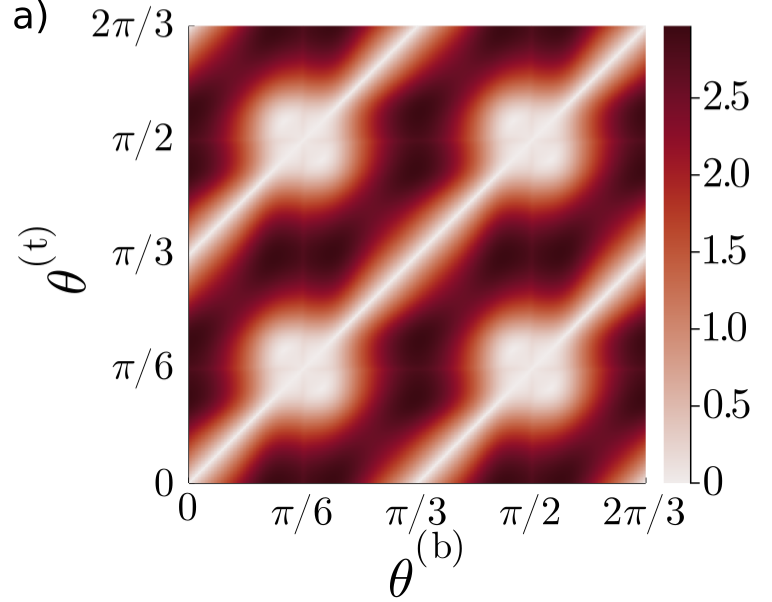}
  \includegraphics[scale=0.20]{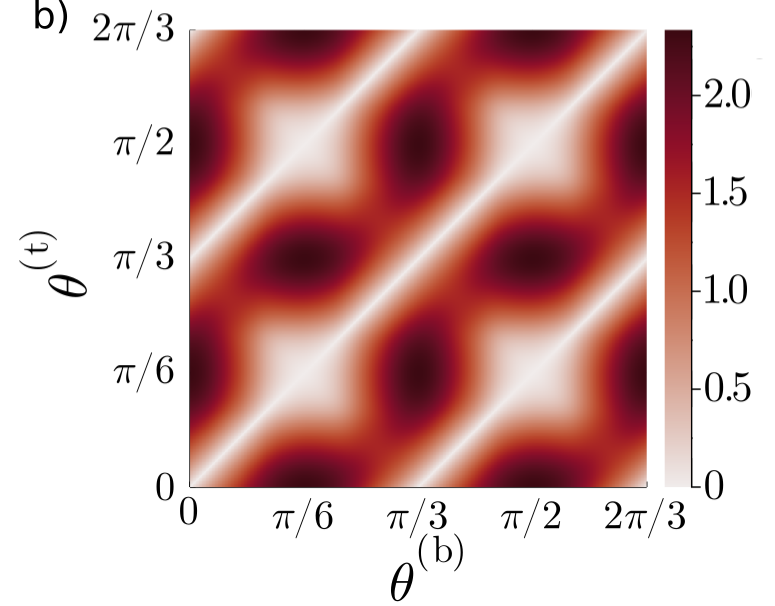}
  \includegraphics[scale=0.21]{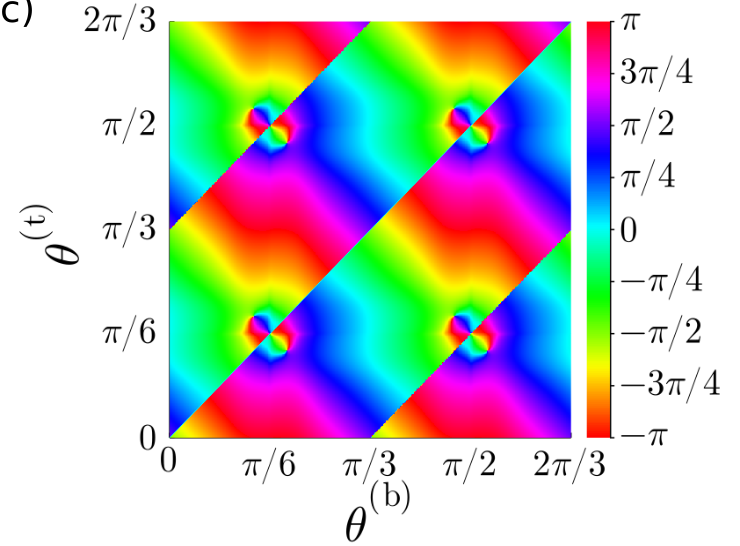}
  \includegraphics[scale=0.21]{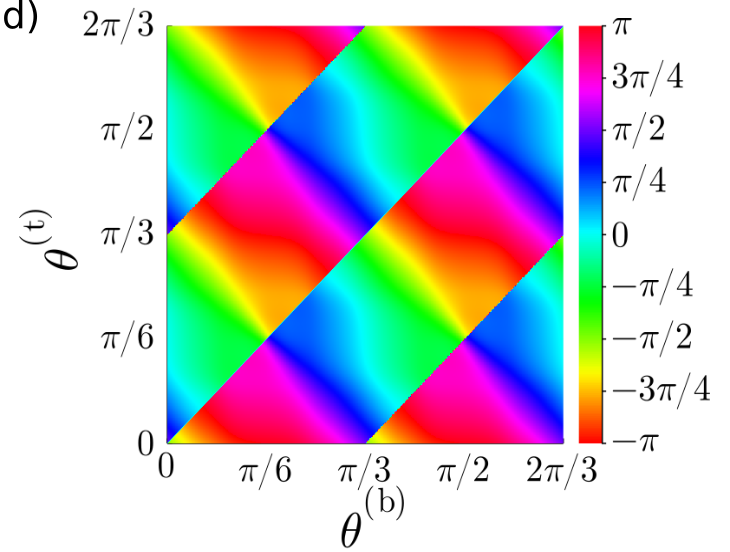}
  \includegraphics[scale=0.20]{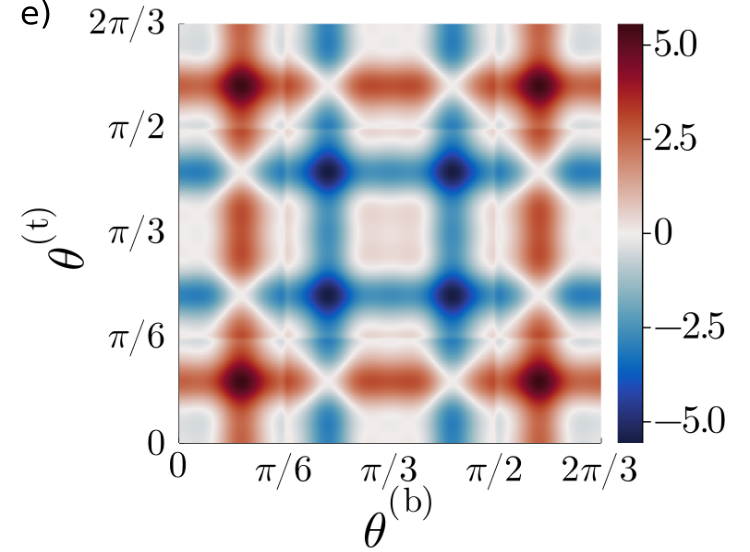}
  \includegraphics[scale=0.20]{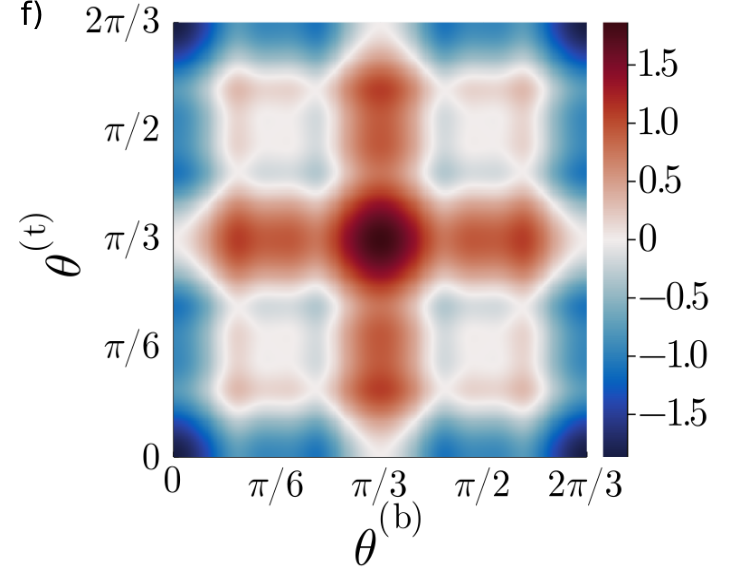}
 \caption{ a) and b) $\lambda_{R}^{(tls)}$ for MoS$_2$ and WSe$_2$ two-sided encapsulation. 
 c) and d) the angle $\vartheta^{(tls)}$ for  MoS$_2$ and WSe$_2$ two-sided encapsulation. e) and f)  $\lambda_{vZ}^{(tls)}$ for MoS$_2$ and WSe$_2$ two-sided encapsulation.
 We used parameters from DFT band structure calculations, see \cite{calculation_parameters_note}.
  \label{fig:trilayer-SOC}}
\end{figure}
%This is our second important result.
Calculations for the twist angle dependence of $\lambda_R$ 
have already been performed in Refs.~\cite{li_twist-angle_2019,alessandro-SOC-and-twist},  therefore we do not show these results here, see Ref.~\cite{supplementary_info} 
for further details.
The results of our numerical calculations for $H_{R}^{(tls)}$ are summarized in Figs.~\ref{fig:trilayer-SOC}(a)-(d). 
Firstly, symmetries that are broken in bilayers can be restored in trilayers for certain  $\theta^{(b)}$ and $\theta^{(t)}$ angles.
If $|\theta^{(b)}-\theta^{(t)}|=l\pi/3$, where  $l$ is  an odd integer, the trilayer stack is inversion symmetric. 
On the other hand, for an even $l$  the stack has a  horizontal mirror plane.
Therefore the induced Rashba SOC must vanish for any integer $l$, as it can be  seen in Figs.~\ref{fig:trilayer-SOC}(a),(b).
For WSe$_2$  encapsulation (Fig.~\ref{fig:trilayer-SOC}(a)) the maximum of $\lambda_{R}^{(tls)}$ is  
found for  $|\theta^{(b)}-\theta^{(t)}|=(2l+1)\pi/6$. %, $l=0,1,2\dots$. 
The maximum value of $\lambda_{R}^{(tls)}$ is around 70\% larger than in the graphene/WSe$_2$ case. Thus,  double-sided encapsulation can significantly enhance the induced Rashba SOC. % and in this respect can be as effective as the application of a strong perpendicular electric field\cite{Gmitra_TMDC-proximity_2016}.
 Surprisingly, when using MoS$_2$ for double encapsulation (Fig.~\ref{fig:trilayer-SOC}(b)), we find that $\lambda_{R}^{(tls)}$ basically cannot be enhanced above 
 the value obtained for one sided proximity effect.
 This can be understood by considering the explicit dependence of $\lambda_{R}^{(t,b)}$ and  $\vartheta_{R}^{(t,b)}$ on $\theta^{(t,b)}$,
 see Ref.~\cite{supplementary_info} for details.  
 The origin of the  extended regions where $\lambda_{R}^{(tls)}$ is very small is due to the fact that for those twist angles $\lambda_R$ and $\vartheta_{R}$ change slowly in both layers and therefore they approximately cancel in Eq.~(\ref{eq:trilayer-complex-lambda_R}).
 In Figs.~\ref{fig:trilayer-SOC}(c,d) we show  the phase $\vartheta^{(tls)}$ of Eq.~(\ref{eq:trilayer-complex-lambda_R}), which determines the winding of the SOC field $\mathbf{s}(\mathbf{q})$ in the case of double encapsulation. 
 The apparent diagonal lines correspond to $\lambda_{R}^{(tls)}=0$ where  $\vartheta^{(tls)}$ is not defined. 
 
For completeness, we also discuss the  twist angle dependence of the induced  valley Zeeman SOC in TMDC/graphene/TMDC trilayers, see Figs.~\ref{fig:trilayer-SOC}(e,f). 
The strength of the valley Zeeman type SOC is simply given by 
$\lambda_{vZ}^{(tls)}=\lambda_{vZ}^{(b)}(\theta^{(b)})+\lambda_{vZ}^{(t)}(\theta^{(t)})$. 
If the two TMDC layers are (nearly) aligned, they can \emph{double} the strength of the induced valley Zeeman SOC.  %For  $|\theta^{(b)}-\theta^{(t)}|=(2l+1)\pi/3$, 
When the whole stack has inversion symmetry, the effect of the two layers cancel and $\lambda_{vZ}^{(tls)}=0$.  The valley Zeeman SOC also vanishes along the lines 
 $\theta^{(b)}+\theta^{(t)}=\pi/3,\pi$, this is a combined effect of time reversal and three-fold rotation symmetry of the TMDC layers. 
We also note that $\lambda_{vZ}^{(tls)}$ depends  sensitively 
on what kind of TMDC is used in the stacks.  The difference between the effects of  WSe$_2$ (Fig.~\ref{fig:trilayer-SOC}(e)) and MoS$_2$(Fig.~\ref{fig:trilayer-SOC}(f)) 
is mainly due to the different alignment of graphene's Dirac point with the TMDC bands. 
The calculations shown in Fig.~\ref{fig:trilayer-SOC}, together with Eq.~(\ref{eq:trilayer-complex-lambda_R}) are the main results of this work.

In the preceding discussions of graphene/TMDC and TMDC/graphene/TMDC heterostructures we have neglected a possible lateral shift of the graphene layer 
 with respect to the TMDCs and  moir\'e effects. Regarding the lateral shift, it does not affect our results\cite{supplementary_info}. 
 Considering the moir\'e effects, in  graphene/TMDC bilayers they are present only at energy scales larger than 2 eV\cite{supplementary_info,pierucci_graphene-MoS2}, i.e., 
 they are negligible for energies close to the Dirac point of graphene. The situation might be different in TMDC/graphene/TMDC trilayers. Based on previous work of  Refs.~\onlinecite{Peeters-supermoire,Koshino-supermoire} on hBN/graphene/hBN trilayers, one may expect that supermoir\'e effects may become important when the TMDC layers are nearly aligned: $0<|\theta^{(t)}-\theta^{(b)}|\lesssim 1^{\circ}$
 (see Ref.~\cite{supplementary_info} for further details).  The discussion of supermoir\'e is  beyond the scope of the present work.

\begin{figure}[ht!]
 \includegraphics[scale=0.24]{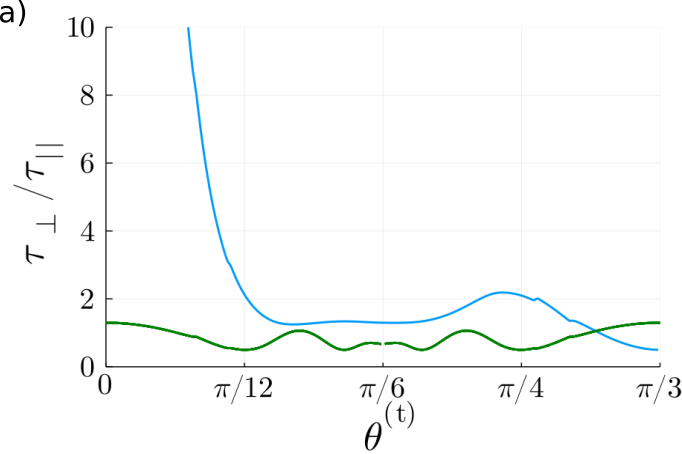}
  \includegraphics[scale=0.2]{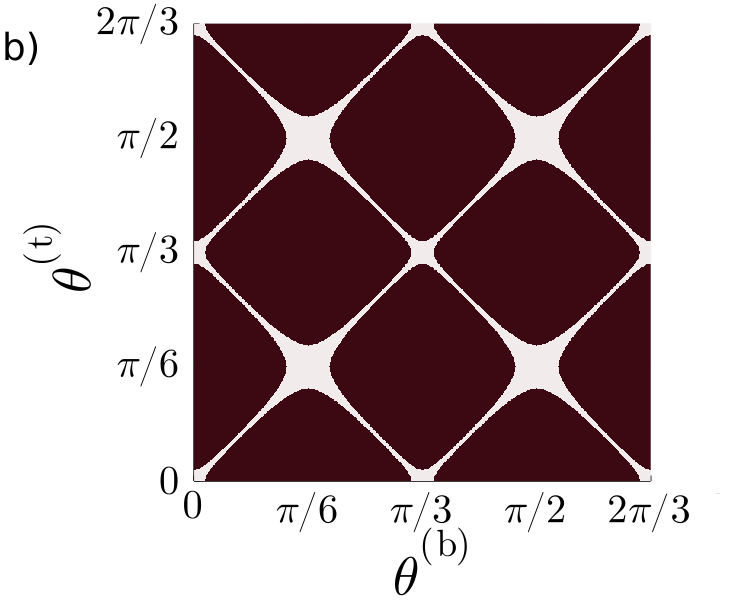}
\includegraphics[scale=0.28]{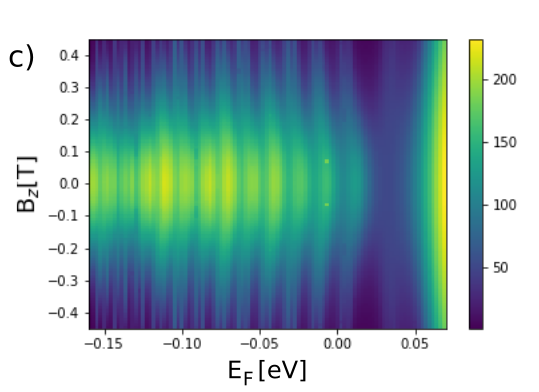}
 \includegraphics[scale=0.28]{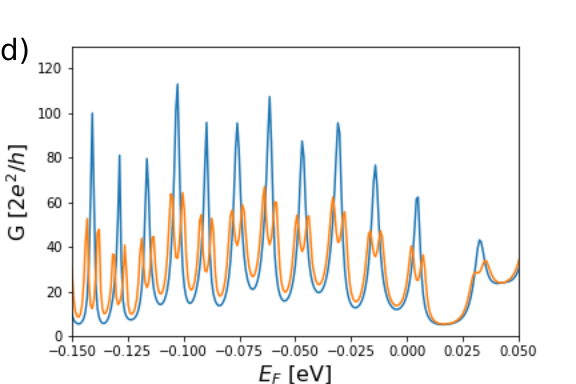}
 \caption{a) Calculated spin lifetime anisotropy in a TMDC/graphene/TMDC heterostructure as a function of $\theta^{(t)}$ for 
 $\theta^{(b)}=0^{\circ}$ (blue) and $\theta^{(b)}=30^{\circ}$ (green).  b) white color indicates the regions in the $(\theta^{(t)}$, $\theta^{(b)})$ space where  the REE is expected to vanish for a WSe$_2$/graphene/WSe$_2$ trilayer. 
 In the calculations for  a) and b) we used DFT parameters\cite{calculation_parameters_note}.
 c)  Conductance $G$ (in  units of $2 e^2/h$) through an $npn$ junction as a function of magnetic field $B_z$ and Fermi energy $E_F$ for $\lambda_{vZ}=2.6$meV. d) 
 Cross section of c) at fixed $B_z=0.436$T. $\lambda_{vZ}=0$meV (blue) and $\lambda_{vZ}=2.6$meV (orange).
 \label{fig:transport-effects}}
\end{figure}

 \textit{Experimental predictions} 
 Several recent works \cite{ghiasi_large_2017,benitez_strong-anistropy-2018,xu_BLG_spinanistropy_2018}  measured an anisotropy of  the  out-of-plane $\tau_{\perp}$ 
 and in-plane $\tau_{\|}$ spin lifetimes in graphene/TMDC heterostructures which can be interpreted as direct proof of induced SOC in graphene. Namely, according to the theoretical calculations of  Ref\cite{cummings_anisotropy_2017}, if there is a strong intervalley scattering in graphene, then the ratio of the spin lifetimes is given by  
 $
 \tau_{\perp}/\tau_{\|}=(\lambda_{vZ}/\lambda_{R})^2 (\tau_{iv}/\tau_{p})+1/2
 $, where $\tau_{iv}$ ($\tau_{p}$) is the intervalley (momentum) scattering time.
(For ultraclean samples with SOC comparable to or larger than the disorder-induced quasiparticle broadening, a qualitatively different spin relaxation anisotropy is derived in Ref.~\cite{offidani_microscopic_2018}.)
  Let us consider a WSe$_2$/graphene/WSe$_2$ heterostructure and assume that $\theta^{(b)}$ is kept fixed while $\theta^{(t)}$ is changed. 
Note,  that the ratio  $\lambda_{vZ}^{}/\lambda_{R}^{}$ can be tuned in a  wider range in  trilayer structures than in bilayers.  
For example, $\lambda_{R}^{}$  is never zero for graphene/TMDC 
whereas one can choose $\theta^{(b)}$ and $\theta^{(t)}$ such that  $\lambda_{vZ}^{(tls)}\neq 0$ and $\lambda_{R}^{(tls)}= 0$, see Figs.~\ref{fig:trilayer-SOC}(a),(e).
Using  $\tau_{iv}/\tau_{p}\approx 5$ as in Ref\cite{cummings_anisotropy_2017}, 
we plot $\tau_{\perp}/\tau_{\|}$ as a function of $\theta^{(t)}$ for  $\theta^{(b)}=0^{\circ}$  and $\theta^{(b)}=30^{\circ}$ in Fig.~\ref{fig:transport-effects}(a). 
When $\theta^{(b)}=0$ and $1^{\circ}<\theta^{(t)}\lesssim 15^{\circ}$ then $\lambda_R^{(tls)}$ becomes small but $\lambda_{vZ}^{(tls)}$ is finite, 
%(see Figs.~\ref{fig:trilayer-SOC}(a) and (e)), 
therefore $\tau_{\perp}/\tau_{\|}$   strongly increases as a function of  $\theta^{(t)}$. This enhancement happens before supermoir\'e effects might become important 
for $\theta^{(t)}\lesssim 1^{\circ}$.
In contrast, if  $\theta^{(b)}=30^{\circ}$ then $\tau_{\perp}/\tau_{\|}$ remains finite
for all $\theta^{(t)}$ angles.  This  dramatic difference in the $\theta^{(t)}$ dependence of  $\tau_{\perp}/\tau_{\|}$ is clearly a consequence of the wider 
tunability of $\lambda_{vZ}^{(tls)}/\lambda_{R}^{(tls)}$ in trilayers. % and we expect that it should  be observable in experiments. 

One can  expect  that  charge-to-spin conversion measurements should also be affected by the interlayer twist. 
Let us consider a graphene/TMDC bilayer and for simplicity assume that i) $\lambda_{vZ}\ll\lambda_R$ and ii) a dc electric field is applied along the $\hat{x}$ direction.
In steady state the $\hat{y}$ component of the spin density is given by
$
\langle S_y\rangle=\frac{1}{2} \int \frac{d^2 \mathbf{q}}{4\pi}  s_y(\mathbf{q})\delta f_{\mathbf{q}}%\propto \cos(\vartheta_{tot}-\theta), 
$
where $\delta f_{\mathbf{q}}$ is the deviation of the quasiparticle distribution with respect to equilibrium. 
Using  the semiclassical argument given in Ref.\cite{ferreira-charge_to_spin2017}, for Fermi energies close to the Dirac point $E_D$ such that $|E_F-E_D|< \lambda_R(\theta)$, 
i.e., when there is a single  Fermi surface,  one finds that $\langle S_y\rangle \propto \cos(\vartheta_{R}+\theta)$. This suggests that the 
Rashba-Edelstein effect (REE) may  vanish (when $\vartheta_{R}+\theta=(2 l+1)\pi/2$) even though $\lambda_R(\theta)$ is 
not required to be zero  by symmetry for any $\theta$.  $\langle S_y\rangle$ can also change sign as a function of the interlayer twist angle, 
because $\vartheta_{R}+\theta\in[0,2\pi]$, see Figs.~\ref{fig:rashba-phase}(b),(c). 
Regarding the more  realistic situation  when  there are two spin-polarized Fermi surfaces in graphene for $|E_F-E_D|>\lambda_R$, 
looking at Figs.~\ref{fig:rashba-phase}(g) one can understand 
that   $\langle S_y\rangle$  can be zero as a function of twist angle in this case as well.
Furthermore, in twisted TMDC/graphene/TMDC trilayers, the REE %Rashba-Edelstein effect
can vanish either because $\lambda_R^{(tls)}=0$ or because $\vartheta^{(tls)}=(2l+1)\pi/2$, 
which can be expressed compactly as $\textrm{Re}\left(\bar{\lambda}_R^{(tls)}\right)=0$. 
As an example, the region where this condition is fulfilled in the $(\theta^{(t)}$, $\theta^{(b)})$ space 
for  WSe$_2$/graphene/WSe$_2$ trilayers is shown in Fig.~\ref{fig:transport-effects}(b).
%Interestingly, the longitudinal component  $\langle S_x\rangle \propto \sin(\vartheta_{tot}-\theta)$ is expected to be non-zero as well.

While  the spin-lifetime anisotropy measurements require diffusive samples, in Figs.~\ref{fig:transport-effects} (c)-(d) we show an example of how strong SOC 
can affect ballistic transport properties.  Highly transparent $np$ and $npn$ junctions in graphene have recently been realized in several experiments\cite{Young-electronoptics2009,Rickhaus2013,Rickhaus2015,CoryDean-electronoptics2016,MakkP-pnjunction2017,Dauber2020} 
and Fabry-Perrot type interference
measurements\cite{MakkP-pnjunction2017,Dauber2020}. Assume  now that  a TMDC/graphene/TMDC trilayer is tuned with interlayer twist angles 
such that the Rashba SOC is switched-off and simultaneously the valley Zeeman  SOC enhanced. We calculate the conductance\cite{supplementary_info} through a smooth  
graphene $npn$ junction\cite{Shytov_npn} as a function of out-of-plane magnetic field $B_z$ and Fermi energy $E_F$ for $\lambda_{vZ}=2.6$meV,
see Fig.~\ref{fig:transport-effects}(c).  One can observe that for $|B_z|>0.3$T  the high-conductance ridges are split. This is  apparent in Fig.~\ref{fig:transport-effects}(d), 
where we compare the cases when $\lambda_{vZ}=0$  (blue) and   $\lambda_{vZ}=2.6$meV (orange) for a fixed magnetic field $B_z=0.436$T.  
Our calculations also indicate that the Rashba type SOC does not have a similar effect on the conductance ridges (not shown). 
An additional experimental probe of the twist angle dependent SOC may be the measurement of the reflection of electrons at a planar junction, which is briefly discussed 
in Ref.~\cite{supplementary_info}. 

\textit{Summary} We found that 
the induced Rashba type SOC in twisted hexagonal bilayers can be parameterized by the strength 
$\lambda_R$ and a spin-rotation angle $\vartheta_{R}$. This latter can lead to interference effects for $\lambda_R$ in trilayer heterostructures. 
We also calculated  the  valley Zeeman SOC in  twisted  TMDC/graphene/TMDC trilayers. 
Finally, we   discussed  how the interlayer twist angle dependence of the induced SOC  can be deduced  from spin-lifetime anisotropy, charge-to-spin conversion 
and magnetrotransport measurements.

In this work we have neglected  possible lattice relaxation in vdW heterostructures\cite{vdW-relaxation-Nicolini}. 
An important future direction would be to study its effect on the results presented here.

\textit{Acknowledgments}  
We  acknowledge helpful conversations with Péter Makk and Jaroslav Fabian.
This research was supported by the Ministry of Innovation and
Technology and the National Research, Development and Innovation
Office (NKFIH) within the Quantum Information National Laboratory of Hungary.
P.R. and A.K.  were also supported by 
%Hungarian Ministry of Human Capacities, and NKFIH through the Quantum Technology National Excellence Program %(No.2017-1.2.1-NKP-2017-00001), 
the ELTE Institutional Excellence Program (TKP2020-IKA-05), and the
Hungarian Scientific Research Fund (OTKA) Grants  No.K134437 and  No.NN127903 (Topograph FlagERA project).
A.K. acknowledges support from the Hungarian Academy of Sciences through the Bolyai János Stipendium (BO/00603/20/11) as well.

\textit{Note added}
During the preparation of this manuscript we have become aware of 
Ref.~\cite{Naimer2021twistangle}, where the authors discuss the twist
angle dependence of the induced SOC in graphene/TMDC structures using 
DFT calculations.

\bibliography{bibliography}

\begin{thebibliography}{75}
\bibitem [{cal()}]{calculation_parameters_note}%
  \BibitemOpen
  \href@noop {} {\bibinfo  {journal} {We describe the energy of the Dirac point
  of graphene in the band gap of the TMDC by a number $f_G \in [0, 1]$. Its
  value is a linear function of the position of the Dirac point in the TMDC
  band gap. When $f_G = 0$, the Dirac point is aligned with the TMDC valence
  band edge, for $f_G = 1$ the Dirac point has the same energy as the TMDC
  conduction band edge. In the calculations involving MoS$_2$, we used DFT
  parameters $E_g=1.807$eV, $f_G = 0.974$ and experimental
  parameters\cite{pierucci_graphene-MoS2} $E_g=2$eV, $f_G = 0.55$. In the case
  of WSe$_2$, DFT parameters were $E_g=1.638$eV, $f_G = 0.161$ and experimental
  ones\cite{nakamura_graphene-WSe2} $E_g=1.95$eV, $f_G = 0.426$}\ }\BibitemShut
  {NoStop}%
\end{thebibliography}

\end{document}

\title{Supplementary information to the paper ``Quantum interference  tuning of spin-orbit coupling in twisted van der
Waals trilayers''}

\author{Csaba G. P\'eterfalvi}
%  \email{csaba.peterfalvi@uni-konstanz.de}
  \affiliation{Department of Physics, University of Konstanz, D-78464 Konstanz, Germany}
  
  \author{Alessandro David}
  \affiliation{Peter Gr\"unberg Institute - Quantum Control (PGI-8), Forschungszentrum J\"ulich GmbH, J\"ulich, Germany}
  
  \author{P\'eter Rakyta}
  \affiliation{Department of Physics of Complex Systems, E\"otv\"os Lor\'and University, Budapest, Hungary}
  \affiliation{Quantum Information National Laboratory, Hungary}
  
  \author{Guido Burkard}
%  \email{guido.burkard@uni-konstanz.de}
  \affiliation{Department of Physics, University of Konstanz, D-78464 Konstanz, Germany}
  
  \author{Andor Korm\'anyos}
%  \email{andor.kormanyos@ttk.elte.hu}
  \affiliation{Department of Physics of Complex Systems, E\"otv\"os Lor\'and University, Budapest, Hungary}

\maketitle

%\tableofcontents

\section{Symmetry properties of the phase $\vartheta_{R}$  }
\label{sec:vartheta-symmetries}

In order to be self-contained and to set the stage, in  Sec.~\ref{sec:intro-to-vartheta} and \ref{sec:vartheta-calc} we  carefully re-derive   previous
results of Ref.~\cite{david_induced_2019} for the  Hamiltonian $H_{R}^{gr}$, which describes the proximity induced Rashba SOC in graphene. 
We then present  two short, but important discussions.  
First,  in Sec.~\ref{sec:lambdaR-sign} we argue that the sign of the SOC coupling $\lambda_R$ depends  on the layer stacking order and this 
is important in trilayer stacks. Next, in  Sec.~\ref{sec:layer-rotation} we discuss  the  effects  of the interlayer twist  on $H_{R}^{gr}$.  
Our most important results in this section follow in Sec.~\ref{sec:vartheta-C3v}.  We prove  that the results for the quantum phase $\vartheta_{R}$ simplify
for interlayer twist angles $\theta=0, \pi/6, \pi/3$, where the bilayer stack has $C_{3v}$ symmetry. 
Finally, in Sec.~\ref{sec:theta-discuss} we give a further  discussion of the results obtained in Sec.~\ref{sec:vartheta-C3v}.

%%%%%%%%%%%%%%%%%%%%%%%%%%%%%%%%%%%%%%%%%%%%%%%%%%%%%%%%%%%%%%%%%%%%%%%%%%%%%%%%%%%%%%%%%%%%%%%%%%%%%%%%%%%%%%%%%%%%%%%%%%%%%%%%%%%%%%%%%%%%%%%%%%%%%%%%

\subsection{Preliminaries}
\label{sec:intro-to-vartheta}

\subsubsection{Lattice vectors}
\label{sec:lattice}

Both graphene and the  TMDC layer  have  hexagonal lattices.  We define the lattice vectors
$\mathbf{a}_1^{}=\frac{a_{}}{2}\left(1, \sqrt{3}\right)^{T}$, 
$\mathbf{a}_2^{}=\frac{a_{}}{2}\left(1, -\sqrt{3}\right))^{T}$ 
(see Fig.~\ref{fig:geom-and-rotated-BZs}(a)) and the corresponding primitive reciprocal lattice vectors 
$\mathbf{b}_1^{}=\frac{2\pi}{a}\left(1, 1/\sqrt{3}\right)^{T}$, 
$\mathbf{b}_2^{}=\frac{2\pi}{a}\left(1, -1/\sqrt{3}\right))^{T}$. 
Here $a=a_{gr}$ ($a=a_{tmdc}$) is the lattice constant for graphene (TMDC). We will distinguish the reciprocal lattice vectors $\mathbf{b}_{1,2}$  
of graphene from  the TMDC ones  by using the notation $\mathbf{b}'_{1,2}$ for the latter. 
The position of the $A$ and $B$ sublattice in the unit cell is given by 
$
\mathbf{t}_{A}^{} =\frac{a_{gr}}{2}
 \left(1, 1/\sqrt{3}\right)^{T}
 $,
$
\mathbf{t}_{B}^{} =\frac{a_{gr}}{2}
 \left(1, -1/\sqrt{3} \right)^{T}.
$
The metal (chalcogen) atoms occupy a position corresponding to the $A$ ($B$) sublattice in the  unit cell of the TMDC.

%%%%%%%%%%%%%%%%%%%%%%%%%%%%%%%%%%%%%%%%%%%%%%%%%%%%%%%%%%%%%%%%%%%%%%%%%%%%%%%%%%%%%%%%%%%%%%%%%%%%%%%%%%%%%%%%%%%%%%%%%%%%%%%%%%%%%%%%%%%%%%%

\subsubsection{Formalism to calculate the induced SOC}
\label{sec:Rashba-SOC-formalism}

As it was shown in Ref\cite{david_induced_2019}, an important contribution to the proximity induced Bychkov-Rashba type SOC in graphene 
comes from virtual interlayer tunneling processes to the TMDC layer. In third order perturbation theory it is given by 
 \begin{equation}
     (H^{gr}_R)_{Xs,X's'} = 
     \sum_{j,b,b',s'',s'''}
      \frac{\left(T(\tau\vect{k}'_j)\right)_{Xs,bs''}
      (H_\text{soc})_{bs'',b's'''}
        \left(T^\dagger(\tau\vect{k}'_j)\right)_{b's''',X's'}}
        {[E_D^\text{gr} - E_{b}^\text{tmdc}(\tau\vect{k}'_j)][E_D^\text{gr} - E_{b'}^\text{tmdc}(\tau\vect{k}'_j)]}.
        \label{eq:Rashba-gen-appendix}
  \end{equation}
  Here $X=\{A,B\}$ runs over sublattice indices of graphene, $s$, $s'$ are spin indices and  $b \neq b'$ are band indices running over the bands of the (isolated) TMDC layer. 
  Spin-flip process can be facilitated by certain off-diagonal elements of the intrinsic SOC matrix of the monolayer TMDC  denoted
  by $(H_\text{soc})_{bs, b's'}^{}$ 
  The spin-flip off-diagonal matrix elements are allowed between pairs of bands $b$, $b'$ if one of the bands is symmetric (even) and the other one is antisymmetric (odd) 
  with respect to reflection on  the horizontal mirror plane of the TMDC (see, e.g., Ref.~\cite{kormanyos_spin-orbit_2014} for further discussion of the SOC in monolayer TMDCs). 
  
\begin{figure}
\begin{center}
\includegraphics[scale=0.45]{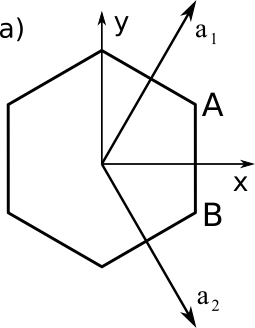}\hspace{1cm}
 \includegraphics[scale=0.60]{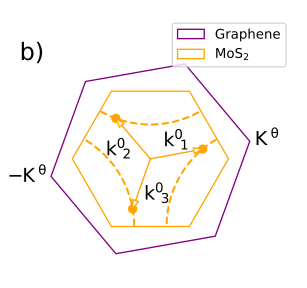}\hspace{1cm}
 \includegraphics[scale=0.45]{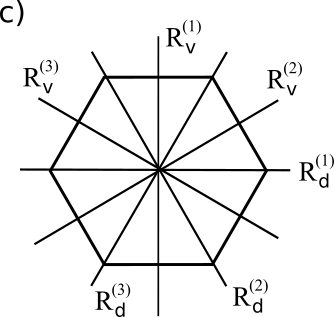}\hspace{1cm}
 \caption{a)  Real space orientation of the hexagonal lattice and the primitive lattice vectors $\mathbf{a}_{1}^{}$, $\mathbf{a}_{2}^{}$.
  b) The rotated BZ and Dirac points $\vect{K}^{\theta}$ of graphene, along with the wavenumbers  $\vect{k}'_j$ in the BZ of the TMDC. 
  As $\theta$  changes in the range $[0..\pi/3]$, the $\vect{k}'_j$  trace out  the arcs shown by dashed lines. c) Reflection planes of the hexagonal BZ. 
  $\mathcal{R}_{v}^{(1,2,3)}$ is relevant for both graphene and the TMDC, $\mathcal{R}_{d}^{(1,2,3)}$ only for graphene.}
 \label{fig:geom-and-rotated-BZs}
 \end{center}
\end{figure}

  We assume that the graphene layer is rotated with respect to the TMDC layer by a twist angle $\theta$. Therefore, as indicated in Fig.~\ref{fig:geom-and-rotated-BZs}(b), the 
  BZ of graphene is also rotated.  For each interlayer twist angle $\theta$ the right-hand side of Eq.~(\ref{eq:Rashba-gen-appendix}) needs to be evaluated for 
  wavenumbers $\tau\vect{k}'_j$ %($j=1,2,3$)  %, $\tau=\pm 1$)  
  in the TMDC Brillouin zone (see Fig.~\ref{fig:geom-and-rotated-BZs}(b))  that   are determined by quasimomentum conservation condition
 $ \tau\vect{K}^{\theta}+\vect{G}^{\theta}=\vect{k}'+\vect{G}'$, 
  such that $|\tau\vect{K}^{\theta}+\vect{G}^{\theta}|=|\vect{K}^{\theta}|$.
 Here  $\tau\vect{K}^{\theta}$, $\tau=\pm 1$ denote the wavevector to the Dirac point of graphene and 
 $\vect{G}^{\theta}$ and ($\vect{G}'$) are reciprocal lattice vectors of  graphene (TMDC).  
 As indicated in Fig.~\ref{fig:geom-and-rotated-BZs}(b), for each $\tau\vect{K}^{\theta}$ there are three wavevectors $\tau\vect{k}'_j$ %($j=1,2,3$) 
 and corresponding reciprocal lattice vectors  $\vect{G}^{\theta}_j$ and $\vect{G}'_j$ that satisfy
 the quasimomentum conservation condition.  These $\tau\vect{k}'_j$ vectors are related by $2\pi/3$ rotations. 
 Explicitly, one finds $\vect{G}^{\theta}_{1,2,3}=-\mathbf{b}_2^{\theta},-\mathbf{b}_1^{\theta},0$ and 
 $\vect{G}'_{1,2,3}=-\mathbf{b}'_2, -(\mathbf{b}'_1-\mathbf{b}'_2), \mathbf{b}'_1$.  
$E_{b}^\text{tmdc}(\tau\vect{k}'_j)$ denote  TMDC bands energies and $E_D^\text{gr}$ is the energy of the Dirac point of graphene.  
  $(T_{\tau\vect{k}'_j})_{Xs,bs''}$ is the tunneling matrix between Bloch states at  $\tau \vect{K}^{\theta}$ to states at  $\tau\vect{k}'_j$. 
  We assume spin-independent tunneling and therefore $(T_{\tau\vect{k}'_j})_{Xs,bs''}$ is diagonal in the spin space. 
 % Since  the Bloch wavefunctions of graphene on the $A$ and $B$ sublattice differ by a  phase, 
  $(T_{\tau\vect{k}'_j})_{Xs,bs''}$ can be written as 
\begin{eqnarray}
\left(T(\tau \bm{k}'_j)\right)_{Xb}= \langle\Psi^{(X)}_{gr}({\mathbf{K}}_1^{\theta},\mathbf{r})|H_{orb}|\Psi^{(b)}_{tmdc}(\mathbf{k}'_j,\mathbf{r}) \rangle =
e^{i\tau \bm{G}^\theta_j \cdot \bm{d}_0} t_b(\tau\vect{k}'_j)  e^{i\tau \phi_j^{X}}
\label{eq:tunneling-matrix}
\end{eqnarray}
Here $H_{orb}$ is the orbital part of the Hamiltonian of the system (without the SOC), 
$\Psi^{(X)}_{gr}({\mathbf{K}}_1^{\theta},\mathbf{r})$ and $\Psi^{(b)}_{tmdc}(\mathbf{k}'_j,\mathbf{r})$ are Bloch wavefunctions of graphene and of band $b$ of 
the TMDC layer, respectively. A possible lateral shift of the graphene lattice with respect to the TMDC lattice is denoted by $\mathbf{d}_0$.
%We take, for simplicity, $\mathbf{r}_0=0$ because this choice  does not affect our results. 
The phase $\phi_j^{X}$ is defined as $\phi_j^{X}=\boldsymbol{t}_{X}\cdot\vect{G}_j$.
The tunneling amplitude $t_b(\tau\vect{k}'_j)$  to band $b$ of the TMDC can be written as 
\begin{equation}
t_b(\tau\mathbf{k}'_j)=e^{-i\tau \bm{G}'_j \cdot \bm{t}'_{X}} t_b^p\left(\tau\mathbf{k}'_j\right)  
+ e^{-i\tau \bm{G}'_j \cdot \bm{t}'_{X'}} t^d_b\left(\tau\mathbf{k}'_j\right),
\label{eq:tunneling-amplitude-def}
\end{equation}
where $\bm{t}'_{X}$ ($\bm{t}'_{X'}$), $X\neq X'$ is the vector pointing to the chalcogen (metal) atom position in the unit cell of the TMDC layer, and  
$t^p_b(\tau\mathbf{k}'_j)$  ($ t^d_b((\tau\mathbf{k}'_j)$) describes  the interlayer tunneling between graphene and the $p$ orbitals of the 
chalcogen ($d$ orbitals of the metal) atoms. 
The calculation of the amplitude $t_b^{p,d}(\tau\vect{k}'_1)$ in terms of materials dependent parameters will be discussed  in Sec.~\ref{sec:tunneling-coeffs}. 

$(H_\text{soc})_{bs,b's'}$ denote the matrix elements of the spin-orbit coupling operator $\hat{H}_{soc}$ between Bloch wavefunctions corresponding to bands 
$b$ and $b'$ of the monolayer TMDC.  For our purposes the matrix elements  between an even $\Psi^{(e)}_{tmdc}(\tau\mathbf{k}'_j,\mathbf{r})$ %$|e, \tau\mathbf{k}'_j\rangle$ 
and an odd  %$|o, \tau\mathbf{k}'_j\rangle$ 
 $\Psi^{(o)}_{tmdc}(\tau\mathbf{k}'_j,\mathbf{r})$
Bloch states are important and they  can be written as
\begin{equation}
 H_{e,o}^{}(\tau\mathbf{k}'_j)=
 \langle \Psi^{(e)}_{tmdc}(\tau\mathbf{k}'_j,\mathbf{r}) |\hat{H}_{soc}|\Psi^{(o)}_{tmdc}(\tau\mathbf{k}'_j,\mathbf{r})\rangle=
 i \lambda_{soc} \left[\alpha_{e,o}^{(x)}(\tau\mathbf{k}'_j) {s}_x +  \alpha_{e,o}^{(y)}(\tau\mathbf{k}'_j) {s}_y\right]. 
 \label{eq:SOC-off-diag}
\end{equation}
Here $\lambda_{soc}$ characterizes the strength of the intrinsic SOC in the TMDC layer. $\hat{H}_{soc}\sim \hat{L}_x \hat{S}_x+\hat{L}_y \hat{S}_y + \hat{L}_z \hat{S}_z= 
\frac{1}{2}(\hat{L}_{+} \hat{S}_{-}+\hat{L}_{-} \hat{S}_{+})+\hat{L}_z \hat{S}_z$, $\hat{L}_{x,y,z}$ are angular momentum operators, 
$\hat{L}_{\pm}=\hat{L}_x\pm i \hat{L}_y$, $\hat{S}_{x,y,z}=\frac{\hbar}{2} s_{x,y,z}$ are spin operators and 
the  Pauli matrices $s_{x,y,z}$ act on the spin degree of freedom. 
The complex numbers $\alpha_{e,o}^{(x,y)}(\tau\mathbf{k}'_j)$ are defined as   
\begin{subequations}
\begin{equation}
i  \, \alpha_{e,o}^{(x)}(\tau\mathbf{k}'_j)=\langle \Psi^{(e)}_{tmdc}(\tau\mathbf{k}'_j,\mathbf{r}) \left| \hat{L}_x \right| \Psi^{(o)}_{tmdc}(\tau\mathbf{k}'_j,\mathbf{r}) \rangle, 
\end{equation}
\begin{equation}
i  \, \alpha_{e,o}^{(x)}(\tau\mathbf{k}'_j)=\langle \Psi^{(e)}_{tmdc}(\tau\mathbf{k}'_j,\mathbf{r}) \left| \hat{L}_y \right| \Psi^{(o)}_{tmdc}(\tau\mathbf{k}'_j,\mathbf{r}) \rangle.
\end{equation}
\label{eq:alphas}
\end{subequations}
 One can make use of the fact that the $\tau \mathbf{k}'_j$ vectors are related by $2\pi/3$ rotations, which is also a symmetry of the TMDC lattice.  
 Following Ref.~\cite{dresselhaus_group_2010} one may write
 $\Psi^{(e (o))}_{tmdc}(\mathcal{R}_{\pm 2\pi/3}\tau\mathbf{k}'_j,\mathbf{r})=\mathcal{R}_{\pm 2\pi/3}\Psi^{(e(o))}_{tmdc}(\tau\mathbf{k}'_j,\mathbf{r}) $
  where $\mathcal{R}_{\pm 2\pi/3}$  denotes rotation by $\pm 2\pi/3$.  Substituting this into Eq.~(\ref{eq:SOC-off-diag}) and taking into account that 
$\mathcal{R}_{2\pi/3}^{} \hat{L}_{\pm} (\mathcal{R}_{2\pi/3}^{})^{\dagger} = e^{\mp i 2 \pi/3} L_{\pm}$, one finds 
 \begin{equation}
 H_{e,o}^{}(\tau\vect{k}'_j)=i \lambda_{soc} \left(
    \begin{array}{cc}
      0 & \alpha_{e,o}^{(-)}(\tau \mathbf{k}'_1) e^{-i (j-1) 2 \pi/3}\\
    \alpha_{e,o}^{(+)}(\tau \mathbf{k}'_1)e^{i (j-1) 2 \pi/3} & 0
    \end{array}
   \right),
   \label{eq:H_eo_j}
\end{equation}
where $\alpha_{e,o}^{(\pm)}(\mathbf{k}'_1)=\alpha_{e,o}^{(x)}(\tau\mathbf{k}'_1)\pm i \alpha_{e,o}^{(y)}(\tau\mathbf{k}'_1)$.

%%%%%%%%%%%%%%%%%%%%%%%%%%%%%%%%%%%%%%%%%%%%%%%%%%%%%%%%%%%%%%%%%%%%%%%%%%%%%%%%%%%%%%%%%%%%%%%%%%%%%%%%%%%%%%%%%%%%%%%%%%%%%%%%%%%%%%%%%%%%%%%%%%%%%%%%% 
 
 \subsection{Derivation of $\vartheta_{R}$ and $\lambda_{R}$}
\label{sec:vartheta-calc}

It is instructive to consider the contribution of a single pair of $b=e$ and $b'=o$  bands in Eq.~(\ref{eq:Rashba-gen-appendix}).
Using Eq.(\ref{eq:tunneling-matrix}), one finds 
 \begin{multline}
      (H^{gr}_R)_{e,o} = \sum_{j} 
      \left(\begin{array}{cc}
       1 & e^{i\tau (\phi_j^A-\phi_j^B)}\\
       e^{-i\tau (\phi_j^A-\phi_j^B)}& 1
      \end{array}\right)
      \otimes
      \frac{\left[{t}_{e,o}^{}(\tau\vect{k}'_j) H_{e,o}^{}(\tau\mathbf{k}'_j)+{t}_{o,e}^{}(\tau\vect{k}'_j)H_{o,e}^{}(\tau\mathbf{k}'_j)\right]}
      {[E_D^\text{gr} - E_{e}^\text{tmdc}(\tau\vect{k}'_j)][E_D^\text{gr} - E_{o}^\text{tmdc}(\tau\vect{k}'_j)]}\\
      = \sum_{j} \mathbf{M}_j(\tau\phi_j^A,\tau\phi_j^B)\otimes
      \frac{\left[{t}_{e,o}^{}(\tau\vect{k}'_j) H_{e,o}^{}(\tau\mathbf{k}'_j)+{t}_{o,e}^{}(\tau\vect{k}'_j)H_{o,e}^{}(\tau\mathbf{k}'_j)\right]}
      {[E_D^\text{gr} - E_{e}^\text{tmdc}(\tau\vect{k}'_j)][E_D^\text{gr} - E_{o}^\text{tmdc}(\tau\vect{k}'_j)]}
      \label{eq:Rashba-gen-eo}
 \end{multline}     
As explained in Sec.~\ref{sec:intro-to-vartheta},  the phases $\phi_j^{A,B}$ in $\mathbf{M}_j(\tau\phi_j^A,\tau\phi_j^B)$ 
are defined as $\phi_j^{A,B}=\boldsymbol{t}_{A,B}\cdot\vect{G}_j$. 
There are a number of possible choices for the  vectors $\boldsymbol{t}_{A,B}$ and for the primitive reciprocal lattice vectors 
$\vect{b}_{1,2}$ of graphene, and the latter  determine $\vect{G}_j$. For a  given  $j$  the phase $\phi_j^{A,B}$ depends on these choices, 
but in general $\phi_j^{A,B} \in \{0, 2\pi/3, -2\pi/3\}$.  The different choices  lead to unitary equivalent 
$\mathbf{M}'_j(\tau\phi_j^A,\tau\phi_j^B)=\mathbf{U}(\tau) \mathbf{M}_j(\tau\phi_j^A,\tau\phi_j^B)\mathbf{U}^{\dagger}(\tau)$ matrices and hence unitary equivalent
$(H^{gr}_R)_{e,o}$ Hamiltonians. 
Importantly, $\mathbf{U}(\tau)$ does not depend  on the geometric twist angle $\theta$. Using Sec.~\ref{sec:lattice}, the $\mathbf{M}_j$ matrices read
\begin{equation}
 \mathbf{M}_j(\tau\phi_j^A,\tau\phi_j^B)=
 \left(
 \begin{array}{cc}
  1 & e^{i \tau \frac{2\pi}{3} j}\\
  e^{-i \tau \frac{2\pi}{3} j} & 1 
 \end{array}
 \right).
 \label{eq:Mj}
\end{equation}
Furthermore, $H_{o,e}^{}(\tau\mathbf{k}'_j)=[H_{e,o}^{}(\tau\mathbf{k}'_j)]^{\dagger}$, 
${t}_{e,o}^{}(\tau\vect{k}'_j)={t}_{e}(\tau\vect{k}'_j) {t}_{o}^{*}(\tau\vect{k}'_j)^{}=(-1)^{\mu}|{t}_{e,o}^{}(\tau\vect{k}'_j)|e^{i\eta(\tau\vect{k}'_j)}$ 
 and  ${t}_{o,e}^{}(\tau\vect{k}'_j)= {t}_{e,o}^{*}(\tau\vect{k}'_j)$. 
 We define $\eta(\tau\vect{k}'_j)={\rm Arg}[{t}_{e,o}^{}(\tau\vect{k}'_j)]\, {\rm mod }\, \pi$ and, 
 as it will be explained in more detail in Sec.~\ref{sec:lambdaR-sign},  the index $\mu=0,1$ gives the sign of the induced Rashba SOC which depends on the layer stacking. 
 Note that the phase factors $e^{i\tau \bm{G}^\theta_j \cdot \bm{d}_0}$, which encode the effect of a lateral shift between the layers in 
 Eq.~(\ref{eq:tunneling-matrix}), have dropped out of the calculations. This means that our  results for the induced Rashba SOC do not depend on such shift. 
 The same conclusion holds for the valley Zeeman SOC as well, which depends on $|t_b(\tau\mathbf{k}'_j)|^2$,  see Ref.~\onlinecite{david_induced_2019}.
 
 Because of the $C_3$ symmetry of the system, ${t}_{e,o}^{}(\tau\vect{k}'_j)$ does not depend on the index $j$ and from  
 time reversal symmetry follows that   $t_{e,o}(-\mathbf{k}'_1)=t_{e,o}^{*}(\mathbf{k}'_1)$. 
 Therefore, by taking into account Eq.~(\ref{eq:H_eo_j}),
\begin{align}
{t}_{e,o}^{}(\tau\mathbf{k}'_j) H_{e,o}^{}(\tau\mathbf{k}'_j)+{t}_{e,o}^{*}(\tau\mathbf{k}'_j)[H_{e,o}^{}(\tau\mathbf{k}'_j)]^{\dagger} 
  = 2 i \lambda_{soc} |{t}_{e,o}^{}(\tau\mathbf{k}'_1)|
  \left(
    \begin{array}{cc}
     0 & \Lambda_{e,o}(\tau\mathbf{k}'_1) e^{-i (j-1) 2 \pi/3}\\
    -\Lambda_{e,o}^{*}(\tau\mathbf{k}'_1) e^{i (j-1) 2 \pi/3} & 0
   \end{array}
  \right),
\label{eq:Lambda_eo_j}
\end{align}
where $\Lambda_{e,o}(\tau\mathbf{k}'_1)=Im[e^{i\eta(\tau\mathbf{k}'_1)} \alpha_{e,o}^{(y)}(\tau\mathbf{k}'_1)]+ i Im[e^{i\eta(\tau\mathbf{k}'_1)} \alpha_{e,o}^{(x)}(\tau\mathbf{k}'_1)]$. 
One can show that $\Lambda_{e,o}(\tau\mathbf{k}'_1)=\tau\Lambda_{e,o}(\mathbf{k}'_1)$ and 
one  may  write $\Lambda_{e,o}^{}(\mathbf{k}'_1)=|\Lambda_{e,o}^{}(\mathbf{k}'_1)|e^{i \tilde{\vartheta}_{e,o}(\mathbf{k}'_1)}$, where 
\begin{equation}
 \tilde{\vartheta}_{e,o}(\mathbf{k}'_1)={\rm Arg}[\Lambda_{e,o}^{}(\mathbf{k}'_1)] =
 {\rm Arctan}\left[\frac{{\rm Im}[e^{i\eta(\mathbf{k}'_1)} \alpha_{e,o}^{(x)}(\mathbf{k}'_1)]}{{\rm Im}[e^{i\eta(\mathbf{k}'_1)} \alpha_{e,o}^{(y)}(\mathbf{k}'_1)]}\right]+m\pi, 
 \hspace{0.5cm}
 m=0,\pm 1, \pm 2\dots
 \label{eq:vartheta-tilde-def}
\end{equation}
Here ${\rm Im}[\dots]$ denotes the imaginary part. 
$ \tilde{\vartheta}_{e,o}(\vect{k}'_1)$ is a quantum phase which depends on  i) the phase $\eta(\mathbf{k}'_1)$ of the interlayer tunneling amplitude between the Bloch states 
of graphene and the TMDC layer and ii) on  off-diagonal matrix elements of the intrinsic SOC of the TMDC layer defined in Eq.(\ref{eq:SOC-off-diag}).

Due to  $C_3$ symmetry,  the denominator of Eq.~(\ref{eq:Rashba-gen-eo}) is the same for all $j$ and because of time reversal symmetry it does 
not depend on $\tau$. Therefore the denominator can be pulled out of the summation. 
Substituting Eqs.~(\ref{eq:Mj}) and (\ref{eq:Lambda_eo_j}) into Eq.~(\ref{eq:Rashba-gen-eo}) 
and  performing  the summation over $j$ one finds
\begin{equation}
(H^{gr}_R)_{e,o}=\lambda_{R,(e,o)}(\vect{k}'_1)
    e^{i\frac{s_z}{2}\vartheta_{e,o}(\vect{k}'_1)}   
\left(
  \begin{array}{cccc}
 0 & 0 & 0 & i \frac{\tau+1}{2}\\
 0 & 0 & -i \frac{\tau-1}{2} & 0 \\
 0 & i \frac{\tau-1}{2} & 0 & 0\\
 -i \frac{\tau+1}{2} & 0 & 0 & 0
\end{array}
\right)      
e^{-i\frac{s_z}{2}\vartheta_{e,o}(\vect{k}'_1)}. 
\label{eq:H_eo-final-big}
\end{equation}
Here  we defined 
\begin{equation}
{\vartheta}_{e,o}(\vect{k}'_1)=\tilde{\vartheta}_{e,o}(\vect{k}'_1)+2\pi/3,
\label{eq:vartheta-def}
\end{equation}
and $\lambda_{R,(e,o)}(\vect{k}'_1)$ is given by\cite{david_induced_2019}
\begin{equation}
 \lambda_{R,(e,o)}(\vect{k}'_1)=\frac{(-1)^{\mu} 6 \lambda_{soc} |{t}_{e,o}^{}(\vect{k}'_1)| |\Lambda_{e,o}(\vect{k}'_1)|}
 {\left[E^\text{gr}_{D}-E_{e}^\text{tmdc}(\vect{k}'_1)\right] 
 \left[E^\text{gr}_{D}-E_{o}^\text{tmdc}(\vect{k}'_1)\right]}.
 \label{eq:lambdaR_eo}
\end{equation}
Using the $\tau_z$  Pauli matrix,  one may rewrite Eq.~(\ref{eq:H_eo-final-big}) in a more compact form:
\begin{equation}
 (H^{gr}_R)_{e,o}= -\frac{\lambda_{R,(e,o)}(\vect{k}'_1)}{2}  e^{i \frac{s_z}{2} {\vartheta}_{e,o}(\vect{k}'_1)} 
  \left(\tau_z \sigma_x s_y + \sigma_y s_x\right) e^{-i \frac{s_z}{2} {{\vartheta}_{e,o}(\vect{k}'_1)}}.
\label{eq:H_R_eo}
\end{equation}
Note, that in the derivation of Eqs.~(\ref{eq:vartheta-def})-(\ref{eq:H_R_eo})  we only made use of  time reversal symmetry and  that the $\vect{k}'_j$ 
vectors are related by $2\pi/3$ rotations.  Therefore these results  are valid for arbitrary interlayer twist angle $\theta$.  
On the other hand,  we considered the contribution of a single pair of $e$ and  $o$ bands of the TMDC to the induced Rashba type SOC. 
As Eq.~(\ref{eq:Rashba-gen-appendix}) shows, one needs to sum over all  $q$ pairs of  $(e,o)$ bands and  one may define the complex Rashba SOC coefficient by
\begin{equation}
  \bar{\lambda}_R =\sum_q
        \lambda_{\text{R},(e,o)_q} e^{i\vartheta_{(e,o)_q}}.
\label{eq:lambdaR-bar}        
\end{equation}
In terms of its magnitude  $\lambda_R(\vect{k}'_1)=|\bar{\lambda}_R(\vect{k}'_1)|$ and phase $\vartheta_{R}(\vect{k}'_1)  = \mathrm{Arg}[\bar{\lambda}_R(\vect{k}'_1)]$,
the Hamiltonian of the induced Rashba SOC can be written as     
\begin{equation}
 H^{gr}_R= -\frac{\lambda_{R}(\vect{k}'_1)}{2}  e^{i \frac{s_z}{2} {\vartheta}_{R}(\vect{k}'_1)} 
  \left(\tau_z \sigma_x s_y + \sigma_y s_x\right) e^{-i \frac{s_z}{2} {{\vartheta}_{R}(\vect{k}'_1)}}.
  \label{eq:H_R-main}
\end{equation}
Note, that both $\lambda_{R}(\vect{k}'_1)$ and ${\vartheta}_{R}(\vect{k}'_1)$ depend implicitly on  $\theta$ through  $\vect{k}'_1$, as shown 
in Fig.~\ref{fig:geom-and-rotated-BZs}. 
We will discuss the  $\theta$ dependence of $\vartheta_{R}$ in more detail in Sec.~\ref{sec:vartheta-C3v}.

%%%%%%%%%%%%%%%%%%%%%%%%%%%%%%%%%%%%%%%%%%%%%%%%%%%%%%%%%%%%%%%%%%%%%%%%%%%%%%%%%%%%%%%%%%%%%%%%%%%%%%%%%%%%%%%%%%%%%%%%%%%%%%%%%%%%%%%%%%%%%%%%%%%%%%%%%%%%%%%%%%%%%%%%%%%%  

\subsection{Staking order dependent sign of $\lambda_R$}
\label{sec:lambdaR-sign}

In our theory the sign of $\lambda_R$ depends on  the stacking order of the graphene and the TMDC layers, %is graphene/TMDC or TMDC/graphene,
i.e., whether is graphene below or above the TMDC layer. 
Such a sign has no physical significance in TMDC/graphene bilayer stacks, but it is important in TMDC/graphene/TMDC trilayers, because it means that the contributions 
of the two TMDC layers to the induced Rashba SOC add up with an opposite sign. 

As explained below  Eq.~(\ref{eq:Mj}),  the induced Rashba SOC is a second order process in interlayer tunneling.  The tunneling amplitude is given by 
${t}_{e,o}^{}(\tau\vect{k}'_j)={t}_{e}(\tau\vect{k}'_j) {t}_{o}^{*}(\tau\vect{k}'_j)^{}$, i.e., it depends on the product of 
the tunneling amplitudes to an $e$ and to an $o$ band of the TMDC.   
$\Psi^{(e)}_{tmdc}(\mathbf{r})$ of an $e$ band of the TMDC is even with respect to the mirror  plane that contains the metal 
atoms of the TMDC layer.  However, the graphene $p_z$ orbitals have a different sign above and below the graphene layer. 
Therefore one can expect  that the matrix element given in Eq.~(\ref{eq:tunneling-matrix}) between the graphene Bloch wavefunction
$\Psi_{gr}(\mathbf{r})$ and $\Psi^{(e)}_{tmdc}(\mathbf{r})$ should have a global sign difference depending on the stacking order of the two layers, i.e., 
whether the graphene layer is above or below the TMDC layer. On the other hand, similar consideration suggests that 
the sign of the amplitude ${t}_{o}(\tau\vect{k}'_j)$  between $\Psi_{gr}(\mathbf{r})$ and $\Psi^{(o)}_{tmdc}(\mathbf{r})$ does not depend on the stacking order.
Therefore, there will be a stacking order dependent sign factor in ${t}_{e,o}^{}(\tau\vect{k}'_j)={t}_{e}(\tau\vect{k}'_j) {t}_{o}^{*}(\tau\vect{k}'_j)^{}$, 
which is described by the index $\mu=0,1$ introduced below Eq.~(\ref{eq:Mj}). 
The stacking order dependence  of  ${t}_{e}(\tau\vect{k}'_j)$ and  ${t}_{o}(\tau\vect{k}'_j)$ can be explicitly shown if  they  are calculated
e.g., using the two-center Slater-Koster parametrization of the corresponding transfer integrals, see Refs.\cite{david_induced_2019,li_twist-angle_2019} for details.

Because of this stacking order dependent sign, the induced Rashba type SOC in TMDC/graphene/TMDC trilayers, which is given by the sum of 
the contributions from the two TMDC layer, reads $H_R^{(tls)}=H_R^{(b)}-H_R^{(t)}$. Taking into account the interlayer twist angle dependence of 
$\lambda_R$ (see Sec.~\ref{sec:bls-SOC-numerics}) and  of $\theta+\vartheta_R$ (shown in Fig.1 of the main text) one can then easily see that $H_R^{(tls)}$ 
vanishes whenever the TMDC/graphene/TMDC stack is either inversion symmetric or has a horizontal mirror symmetry.

%The stacking order dependence  of  ${t}_{e}(\tau\vect{k}'_j)$ and  ${t}_{o}(\tau\vect{k}'_j)$ can be explicitly seen if  they  are calculated 
%using the two-center Slater-Koster parametrization of the corresponding transfer integrals, see Refs.\cite{david_induced_2019,li_twist-angle_2019} for details. 

%%%%%%%%%%%%%%%%%%%%%%%%%%%%%%%%%%%%%%%%%%%%%%%%%%%%%%%%%%%%%%%%%%%%%%%%%%%%%%%%%%%%%%%%%%%%%%%%%%%%%%%%%%%%%%%%%%%%%%%%%%%%%%%%%%%%%%%%%%%%%%%%%%%%%%%%%%%%%%%%%%%%%%%%%%%%  

\subsection{Effects of the interlayer twist}
\label{sec:layer-rotation}

\begin{figure*}[ht!]
\includegraphics[scale=0.32]{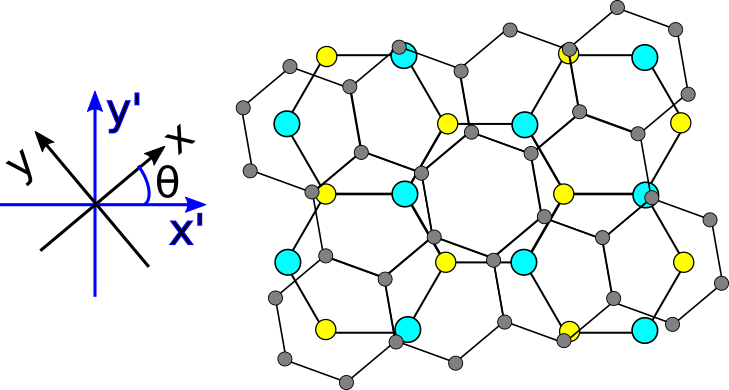}
  \caption{Schematics of a twisted graphene/TMDC bilayer with twist angle $\theta$ and  the coordinate systems $x-y$  and $x'-y'$. 
  \label{fig:coord-sys}}
\end{figure*}

$H^{gr}_R$ in   Eq.~(\ref{eq:H_R-main}) was obtained in the $x'-y'$ coordinate  system of the TMDC layer (see  Fig.~\ref{fig:coord-sys}), where  the  matrix elements of $\hat{H}_{soc}$  were evaluated. 
Our choice of reciprocal lattice vectors leads to $H_{orb}^{gr}=v_F(\tau_z\sigma_x \delta k_x-\sigma_y \delta k_y)$ for the orbital Hamiltonian 
of graphene in the $x-y$ coordinate system, where $\delta\vect{k}=(\delta k_x, \delta k_y)$ is  a small wavevector measured from the $\tau \vect{K}$ point. 
In the $x'-y'$ system  $H_{orb}^{gr}$ is given by 
$H_{orb}^{gr,\theta}=e^{-i\tau_z \frac{\sigma_z}{2}{\theta}}H_{orb}^{gr}e^{i\tau_z \frac{\sigma_z}{2}{\theta}} $. 
The total effective Hamiltonian of graphene is therefore  
\begin{equation}
H_{\rm eff}^{gr,\theta} = e^{-i\tau_z \frac{\sigma_z}{2}{\theta}}H_{orb}^{gr}e^{i\tau_z \frac{\sigma_z}{2}{\theta}}+H^{gr}_R+H^{gr}_{vZ},
\end{equation}
where we have also taken into account the valley-Zeeman  induced SOC, which is described by 
\begin{equation}
H_{vZ}^{gr}=\lambda_{vZ}\tau_z s_z.
\label{eq:HvZ-def}
\end{equation}
$H_{vZ}^{gr}$ is invariant under rotations around the $\hat{z}$ axis and has the same form in both the  $x-y$ and $x'-y'$ coordinate systems.
One may perform a  unitary transformation $e^{i\tau_z \frac{\sigma_z}{2}{\theta}}H_{\rm eff}^{gr,\theta}e^{-i\tau_z \frac{\sigma_z}{2}{\theta}}$. This 
transformation leaves $H_{vZ}$ unchanged and  one finds 
\begin{equation}
% $
H_{\rm eff}^{gr}=H_{orb}^{gr} + e^{i\tau_z \frac{\sigma_z}{2}{\theta}}H_{R}^{gr}e^{-i\tau_z \frac{\sigma_z}{2}{\theta}}+H_{vZ}^{gr}. 
%$.
\label{eq:Heff-total}
\end{equation}
It is convenient to perform another unitary transformation $H_{\rm eff}^{gr}\rightarrow \sigma_x H_{\rm eff}^{gr} \sigma_x$. 
This changes the orbital Hamiltonian 
$
v_F(\tau_z\sigma_x \delta k_x-\sigma_y \delta k_y) \rightarrow v_F(\tau_z\sigma_x \delta k_x+\sigma_y \delta k_y), 
$
leaves $H_{vZ}^{gr}$ unchanged, and transforms  the Rashba Hamiltonian to 
\begin{equation}
\sigma_x e^{i\tau_z \frac{\sigma_z}{2}{\theta}} H_{R}^{gr}  e^{-i\tau_z \frac{\sigma_z}{2}{\theta}}\sigma_x 
= -\frac{\lambda_{R}(\vect{k}'_1)}{2} e^{-i\tau_z \frac{\sigma_z}{2}{\theta}} e^{i \frac{s_z}{2} {\vartheta}_{R}(\vect{k}'_1)}  
  \left(\tau_z \sigma_x s_y - \sigma_y s_x\right)
 e^{-i \frac{s_z}{2} {\vartheta}_{R}(\vect{k}'_1)}  e^{i\tau_z \frac{\sigma_z}{2}{\theta}} .
 \label{eq:H_R-transformed}
\end{equation}
Eq.~(\ref{eq:H_R-transformed}) shows more explicitly the effects of interlayer rotation on the graphene Rashba Hamiltonian 
$
 \tilde{H}^{gr}_R= \frac{\lambda_{R}}{2} \left(\tau_z \sigma_x s_y - \sigma_y s_x\right)
 $
derived previously in Refs.~\cite{kane_quantum-spin-Hall_2005,min_intrinsic_2006}:
i) $\tilde{H}^{gr}_R$  is rotated in real space by $\theta$, an ii)  if $\vartheta_{R}\neq 2 m \pi$, $m$ integer, then
a rotation   in spin-space appears.

%%%%%%%%%%%%%%%%%%%%%%%%%%%%%%%%%%%%%%%%%%%%%%%%%%%%%%%%%%%%%%%%%%%%%%%%%%%%%%%%%%%%%%%%%%%%%%%%%%%%%%%%%%%%%%%%%%%%%%%%%%%%%%%%%%%%%%%%%%%%%%%%%%%%%%%%%%%%%%%%%%

%%%%%%%%%%%%%%%%%%%%%%%%%%%%%%%%%%%%%%%%%%%%%%%%%%%%%%%%%%%%%%%%%%%%%%%%%%%%%%%%%%%%%%%%%%%%%%%%%%%%%%%%%%%%%%%%%%%%%%%%%%%%%%%%%%%%%%%%%%%%%%%%%%%%%%%%%%%%%%%%%%  

 \subsection{Calculation of   $\vartheta_{R}$ for $\theta=0$, $\theta=\pi/6$ and $\theta=\pi/3$  twist angles}
\label{sec:vartheta-C3v} 
  
The calculations of Sec.~\ref{sec:vartheta-calc}  are valid for arbitrary interlayer twist angle $\theta$.   
For $\theta_l=l\pi/6$, $l=0,1,2\dots$ the graphene/TMDC bilayer  has a higher,  $C_{3v}$ symmetry, which implies further constraints on the form of 
${H}^{gr}_R$. By  calculating $\vartheta_{e,o}(\vect{k}'_1)$ defined in Eq.~(\ref{eq:vartheta-def}) for $\theta=0, \pi/6, \pi/3$ explicitly, 
we  show how the results of Sec.~\ref{sec:vartheta-calc} are simplified in this case. %when the bilayer stack has $C_{3v}$ symmetry.  

 \subsubsection{$\theta=0$ twist angle}
 \label{sec:theta0}
 
\begin{figure*}[htb]
\includegraphics[scale=0.4]{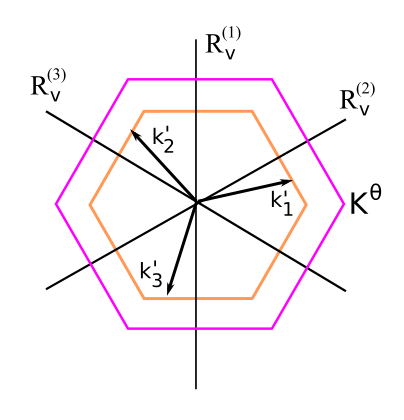}
  \caption{The BZ of graphene (purple) and of the TMDC (orange), along with  $\mathbf{k}'_{1,2,3}$ vectors and the reflection planes 
  $\mathcal{R}_{v}^{1,2,3}$  for $\theta=0$.
  \label{fig:0degkvect}}
\end{figure*}

In Fig.~\ref{fig:0degkvect}  we show the wavevectors $\mathbf{k}'_{1,2,3}$  that satisfy  the quasimomentum conservation as well as the three 
reflection planes $\mathcal{R}_v^{(1,2,3)}$ that the system possess in addition to the $C_3$ symmetry.
First, we will derive a useful relation for the  tunneling amplitude 
$t_{e,o}(\vect{k}'_1)=t_{e}(\vect{k}'_1)t_{o}^{*}(\vect{k}'_1)$, defined below Eq.~(\ref{eq:Mj}).

Using the definition $t_{e (o)}(\tau\vect{k}'_j)$ given in Eq.(\ref{eq:tunneling-matrix}) and denoting $\vect{K}^{\theta=0}=\vect{K}^{(0)}$,   consider the following: 
\begin{align}
t_{e (o)}(\vect{k}'_1) e^{-i\phi^{X}_1} &=\langle\Psi^{(X)}_{gr}({\mathbf{K}}_1^{(0)},\mathbf{r})|H_{orb}|\Psi^{(e(o))}_{tmdc}(\mathbf{k}'_1,\mathbf{r}) \rangle
=\langle (H_{orb} \Psi^{(e (o))}_{tmdc}(\mathbf{k}'_1,\mathbf{r}))^{*}|(\Psi^{(X)}_{gr}({\mathbf{K}}_1^{(0)},\mathbf{r}))^{*}\rangle\nonumber\\
  &=\langle H_{orb} (\Psi^{(e (o))}_{tmdc}(\mathbf{k}'_1,\mathbf{r}))^{*}| (\Psi^{(X)}_{gr}({\mathbf{K}}_1^{(0)},\mathbf{r}))^{*}\rangle 
  =\langle H_{orb} \Psi^{(e (o))}_{tmdc}(-\mathbf{k}'_1,\mathbf{r})| \Psi^{(X)}_{gr}({-\mathbf{K}}_1^{(0)},\mathbf{r})\rangle\nonumber\\ 
  &=\langle H_{orb} \Psi^{(e (o))}_{tmdc}(\mathcal{R}_{v}^{(2)}(\mathbf{k}'_1-\mathbf{b}'_1),\mathbf{r})| \Psi^{(X)}_{gr}(\mathcal{R}_{v}^{(1)} \mathbf{K}_1^{(0)},\mathbf{r})\rangle
  \nonumber\\
  &=\langle H_{orb} \mathcal{R}_{v}^{(2)} \Psi^{(e (o))}_{tmdc}(\mathbf{k}'_1-{\mathbf{b}'_1},\mathbf{r})| \Psi^{(X)}_{gr}(\mathcal{R}_{v}^{(1)} \mathbf{K}_1^{(0)},\mathbf{r})\rangle\nonumber\\
  &= \langle \mathcal{R}_{v}^{(2)} H_{orb}  \Psi^{(e (o))}_{tmdc}(\mathbf{k}'_1-{\mathbf{b}'_1},\mathbf{r})|\Psi^{(X)}_{gr}(\mathcal{R}_{v}^{(1)} \mathbf{K}_1^{(0)},\mathbf{r}) \rangle
  \nonumber\\
  &=\langle H_{orb}  \Psi^{(e (o))}_{tmdc}(\mathbf{k}'_1-{\mathbf{b}'_1},\mathbf{r})| \mathcal{R}_{v}^{(2)}\Psi^{(X)}_{gr}(\mathcal{R}_{v}^{(1)} \mathbf{K}_1^{(0)},\mathbf{r})\rangle\nonumber\\
  &= \langle H_{orb}  \Psi^{(e (o))}_{tmdc}(\mathbf{k}'_1-{\mathbf{b}'_1},\mathbf{r})| \Psi^{(X)}_{gr}(\mathcal{R}_{v}^{(2)}\mathcal{R}_{v}^{(1)} \mathbf{K}_1^{(0)},\mathbf{r})\rangle
  \nonumber\\
  &=\langle H_{orb}  \Psi^{(e (o))}_{tmdc}(\mathbf{k}'_1-{\mathbf{b}'_1},\mathbf{r})| \Psi^{(X)}_{gr}(\mathbf{K}_1^{(0)}-{\mathbf{b}_1^{(0)}},\mathbf{r})\rangle\nonumber\\ 
  &= \left(\langle  \Psi^{(X)}_{gr}(\mathbf{K}_1^{(0)}-{\mathbf{b}_1^{(0)}},\mathbf{r}) | H_{orb} | \Psi^{(e (o))}_{tmdc}(\mathbf{k}'_1-{\mathbf{b}'_1},\mathbf{r})\rangle\right)^{*}
  \nonumber\\
  &=\left(e^{-i\omega^{X}_1}\langle  \Psi^{(X)}_{gr}(\mathbf{K}_1^{(0)},\mathbf{r}) | H_{orb} | \Psi^{(e (o))}_{tmdc}(\mathbf{k}'_1,\mathbf{r})\rangle e^{i\omega^{(e(o))}_1}\right)^{*}
  = t_{e (o)}^{*}(\vect{k}'_1) e^{i\phi^{X}_1} e^{i\omega^{X}_1}e^{-i\omega^{(e(o))}_1}.
\label{eq:tunneling-amp-symmetry}
\end{align}
Here we have made use of i) time reversal symmetry when assuming $H^{*}_{orb}=H_{orb}$, ii) $\mathcal{R}_{v}^{(2)}=(\mathcal{R}_{v}^{(2)})^{-1}$, and iii) $\mathcal{R}_{v}^{(2)}$ 
is a symmetry of the  system therefore $\mathcal{R}_{v}^{(2)} H_{orb} = H_{orb} \mathcal{R}_{v}^{(2)}$.  Moreover, 
 Bloch wavefunctions at equivalent wavenumbers can differ at most by a phase factor, i.e., 
 $|\Psi^{(X)}_{gr}(\mathbf{K}_1^{(0)}-{\mathbf{b}_1^{(0)}},\mathbf{r})\rangle=|\Psi^{(X)}_{gr}(\mathbf{K}_1^{(0)},\mathbf{r})\rangle e^{i\omega^{X}_1}$ and 
 $|\Psi^{(e (o))}_{tmdc}(\mathbf{k}'_1-{\mathbf{b}'_1},\mathbf{r})\rangle=|\Psi^{(e (o))}_{tmdc}(\mathbf{k}'_1,\mathbf{r})\rangle e^{i\omega^{(e(o))}_1}$.
From Eq.~(\ref{eq:tunneling-amp-symmetry}) one finds that 
\begin{equation}
 t_{e,o}(\vect{k}'_1)=t_{e}(\vect{k}'_1)t_{o}^{*}(\vect{k}'_1)%=t_{e}^{*}(\vect{k}'_1)t_{o}^{}(\vect{k}'_1)e^{-i\omega^{(e)}_1}e^{i\omega^{(o)}_1}
 =t_{e,o}^{*}(\vect{k}'_1)e^{-i\omega^{(e)}_1}e^{i\omega^{(o)}_1}.
 \label{eq:teo-symmetry}
\end{equation}

A similar relation can be derived for the matrix elements of $\hat{H}_{soc}$, namely, 
for $\langle\Psi_{}^{(e)}(\mathbf{k}'_1,\mathbf{r})|L_{\pm}|\Psi_{}^{(o)}(\mathbf{k}'_1,\mathbf{r})\rangle$:
\begin{align}
 \langle\Psi_{}^{(e)}(\mathbf{k}'_1,\mathbf{r})|L_{\pm}|\Psi_{}^{(o)}(\mathbf{k}'_1,\mathbf{r})\rangle &= 
  \langle(L_{\pm} \Psi_{}^{(o)}(\mathbf{k}'_1,\mathbf{r}))^{*}|(\Psi_{}^{(e)}(\mathbf{k}'_1,\mathbf{r}))^{*}\rangle 
  = -\langle L_{\mp}\Psi_{}^{(o)}(-\mathbf{k}'_1,\mathbf{r})|\Psi_{}^{(e)}(-\mathbf{k}'_1,\mathbf{r})\rangle\nonumber\\
   &= -\langle L_{\mp} \Psi_{}^{(o)}(\mathcal{R}_{v}^{(2)}(\mathbf{k}'_1-\mathbf{b}'_1),\mathbf{r})|\Psi_{}^{(e)}(\mathcal{R}_{v}^{(2)}(\mathbf{k}'_1-\mathbf{b}'_1),\mathbf{r})
   \rangle\nonumber\\
   &= -\langle L_{\mp} \mathcal{R}_{v}^{(2)} \Psi_{}^{(o)}(\mathbf{k}'_1-\mathbf{b}'_1,\mathbf{r})|\mathcal{R}_{v}^{(2)}\Psi_{}^{(e)}(\mathbf{k}'_1-\mathbf{b}'_1,\mathbf{r})
   \rangle\nonumber\\
   &= -\langle e^{\pm i \frac{2 \pi}{3}} \mathcal{R}_{v}^{(2)} [L_{\pm} \Psi_{}^{(o)}(\mathbf{k}'_1-\mathbf{b}'_1,\mathbf{r})]|
      \mathcal{R}_{v}^{(2)}\Psi_{}^{(e)}(\mathbf{k}'_1-\mathbf{b}'_1, \mathbf{r})\rangle\nonumber\\
   &= -e^{\mp i \frac{2 \pi}{3}} \langle [L_{\pm} \Psi_{}^{(o)}(\mathbf{k}'_1-\mathbf{b}'_1,\mathbf{r})]| \Psi_{}^{(e)}(\mathbf{k}'_1-\mathbf{b}'_1, \mathbf{r})\rangle\nonumber\\
   &= -e^{\mp i \frac{2 \pi}{3}} \left(\langle \Psi_{}^{(e)}(\mathbf{k}'_1-\mathbf{b}'_1,\mathbf{r})| L_{\pm} | \Psi_{}^{(o)}(\mathbf{k}'_1-\mathbf{b}'_1, \mathbf{r})\rangle\right)^{*}
   \nonumber\\
   &= -e^{\mp i \frac{2 \pi}{3}} \left(\langle \Psi_{}^{(e)}(\mathbf{k}'_1,\mathbf{r})| L_{\pm} | \Psi_{}^{(o)}(\mathbf{k}'_1, \mathbf{r})\rangle\right)^{*} 
   e^{i\omega^{(e)}_1}e^{-i\omega^{(o)}_1}.
 \label{eq:SOC-offdiag-0pi}
\end{align}
In terms of $\alpha_{e,o}^{(\pm)}(\mathbf{k}'_1)$, introduced after Eq.~(\ref{eq:H_eo_j}),  Eq.~(\ref{eq:SOC-offdiag-0pi}) means 
that $\alpha_{e,o}^{(\pm)}(\mathbf{k}'_1)=-e^{\mp i\frac{2\pi}{3}}[\alpha_{e,o}^{(\pm)}(\mathbf{k}'_1)]^{*} e^{i\omega^{(e)}_1}e^{-i\omega^{(o)}_1}$.
Therefore, 
\begin{subequations}
\begin{align}
 \alpha_{e,o}^{(x)}({\mathbf{k}}_1)&=\frac{1}{2}\left(\alpha^{(+)}_{e,o}(\mathbf{k}'_1)+\alpha^{(-)}_{e,o}(\mathbf{k}'_1)\right) =  
-\frac{1}{2}\left(e^{- i \frac{2\pi}{3}} [\alpha^{(+)}_{e,o}(\mathbf{k}'_1)]^{*} + e^{i \frac{2\pi}{3}} [\alpha^{(-)}_{e,o}(\mathbf{k}'_1)]^{*}\right)
e^{i\omega^{(e)}_1}e^{-i\omega^{(o)}_1}\nonumber \\
&= -\frac{1}{2}\left(-\frac{1}{2}\left([\alpha^{(+)}_{e,o}(\mathbf{k}'_1)]^{*}+[\alpha^{(-)}_{e,o}(\mathbf{k}'_1)]^{*}\right)- 
i \frac{\sqrt{3}}{2}\left([\alpha^{(+)}_{e,o}(\mathbf{k}'_1)]^{*}-[\alpha^{(-)}_{e,o}(\mathbf{k}'_1)]^{*}\right)  \right) e^{i\omega^{(e)}_1}e^{-i\omega^{(o)}_1},\\
\alpha_{e,o}^{(y)}({\mathbf{k}}_1)&=\frac{1}{2i}\left(\alpha^{(+)}_{e,o}(\mathbf{k}'_1)-\alpha^{(-)}_{e,o}(\mathbf{k}'_1)\right)=
-\frac{1}{2 i}\left(e^{- i \frac{2\pi}{3}}[\alpha^{(+)}_{e,o}(\mathbf{k}'_1)]^{*}-e^{i \frac{2\pi}{3}} [\alpha^{(-)}_{e,o}(\mathbf{k}'_1)]^{*}\right) e^{i\omega^{(e)}_1}e^{-i\omega^{(o)}_1}
\nonumber \\
&= -\frac{1}{2 i}\left(-\frac{1}{2}\left([\alpha^{(+)}_{e,o}(\mathbf{k}'_1)]^{*}-[\alpha^{(-)}_{e,o}(\mathbf{k}'_1)]^{*}\right)- 
i \frac{\sqrt{3}}{2}\left([\alpha^{(+)}_{e,o}(\mathbf{k}'_1)]^{*}+[\alpha^{(-)}_{e,o}(\mathbf{k}'_1)]^{*}\right)  \right) e^{i\omega^{(e)}_1}e^{-i\omega^{(o)}_1}.
\end{align}
\label{eq:alphaxy-0pi}
\end{subequations}
Eqs.~(\ref{eq:alphaxy-0pi}) can be written as 
\begin{equation}
 \left(
 \begin{array}{c}
  \alpha_{e,o}^{(x)}(\mathbf{k}'_1)  \\
  \alpha_{e,o}^{(y)}(\mathbf{k}'_1)  
 \end{array}
 \right) = -\frac{1}{2}
 \left(
 \begin{array}{cc}
  1 &   \sqrt{3} \\
  \sqrt{3} & -1
 \end{array}
 \right)
  \left(
  \begin{array}{c}
     \left(\alpha_{e,o}^{(x)}(\mathbf{k}'_1)\right)^{*} \\
     \left(\alpha_{e,o}^{(y)}(\mathbf{k}'_1)\right)^{*}
  \end{array}
   \right)
   e^{i\omega^{(e)}_1}e^{-i\omega^{(o)}_1},
  \label{eq:alphaxy-0pi-matrix}
\end{equation}

One now calculate ${\vartheta}_{e,o}(\vect{k}'_1)$ for $\theta=0$. First, 
 combining  Eq.~(\ref{eq:teo-symmetry}) and Eq.~(\ref{eq:alphaxy-0pi-matrix}) one  arrives at 
\begin{equation} 
t_{e,o}(\vect{k}'_1)
 \left(
 \begin{array}{c}
  \alpha_{e,o}^{(x)}(\mathbf{k}'_1)  \\
  \alpha_{e,o}^{(y)}(\mathbf{k}'_1)  
 \end{array}
 \right) = -\frac{1}{2}
 \left(
 \begin{array}{cc}
  1 &   \sqrt{3} \\
  \sqrt{3} & -1
 \end{array}
 \right)
  \left(
  \begin{array}{c}
     \left(\alpha_{e,o}^{(x)}(\mathbf{k}'_1)\right)^{*} \\
     \left(\alpha_{e,o}^{(y)}(\mathbf{k}'_1)\right)^{*}
  \end{array}
   \right) t_{e,o}^{*}(\vect{k}'_1).
  \label{eq:alphaxy-teto-0pi}
\end{equation}
This means that 
%\begin{equation}
$
 {\rm Im}\left[t_{e,o}(\vect{k}'_1) \alpha_{e,o}^{(x)}(\mathbf{k}'_1)\right]=-\frac{1}{2} {\rm Im}\left[[t_{e,o}(\vect{k}'_1) \alpha_{e,o}^{(x)}(\mathbf{k}'_1)]^{*}\right]
 -\frac{\sqrt{3}}{2}   {\rm Im}\left[[t_{e,o}(\vect{k}'_1) \alpha_{e,o}^{(y)}(\mathbf{k}'_1)]^{*}\right],
 $
%\end{equation}
i.e., ${\rm Im}\left[t_{e,o}(\vect{k}'_1) \alpha_{e,o}^{(x)}(\mathbf{k}'_1)\right]=
\sqrt{3}\,{\rm Im}\left[t_{e,o}(\vect{k}'_1) \alpha_{e,o}^{(y)}(\mathbf{k}'_1)\right].
$
Regarding Eq.~(\ref{eq:vartheta-tilde-def}),   it follows that 
\begin{equation}
 {\rm Arctan}\left[\frac{{\rm Im}[e^{i\eta(\mathbf{k}'_1)} \alpha_{e,o}^{(x)}(\mathbf{k}'_1)]}{{\rm Im}[e^{i\eta(\mathbf{k}'_1)} \alpha_{e,o}^{(y)}(\mathbf{k}'_1)]}\right]
 ={\rm Arctan}\left[\frac{{\rm Im}[t_{e,o}(\mathbf{k}'_1) \alpha_{e,o}^{(x)}(\mathbf{k}'_1)]}{{\rm Im}[t_{e,o}(\mathbf{k}'_1) \alpha_{e,o}^{(y)}(\mathbf{k}'_1)]}\right]=
 \frac{\pi}{3},
\end{equation}
and using  Eq.~(\ref{eq:vartheta-def}) one finds  ${\vartheta}_{e,o}(\mathbf{k}'_1)=(m+1)\pi$. Therefore, one finds from Eq.~(\ref{eq:H_R-transformed}) that for $\theta=0$
%\begin{equation}
$ H_{R}^{gr}= (-1)^{m}\frac{\lambda_R(\mathbf{k}'_1)}{2}\left(\tau_z \sigma_x s_y - \sigma_y s_x\right).$
%\end{equation}

%%%%%%%%%%%%%%%%%%%%%%%%%%%%%%%%%%%%%%%%%%%%%%%%%%%%%%%%%%%%%%%%%%%%%%%%%%%%%%%%%%%%%%%%%%%%%%%%%%%%%%%%%%%%%%%%%%%%%%%%%%%%%%%%%%%%%%%%%%%%%%%%%%%%%%%%%%%%%%%%%%%%%%%%%%%%  

\subsubsection{$\theta=\pi/6$ twist angle}
 \label{sec:theta-pi6}

Graphene has two, inequivalent sets of vertical  reflection planes, see Fig.~\ref{fig:geom-and-rotated-BZs}. 
For $\theta=\pi/6$, the $\mathcal{R}_d^{(1,2,3)}$ set aligns with the TMDC's $\mathcal{R}_v^{(1,2,3)}$ reflection planes, therefore 
the stack again has  $C_{3v}$ symmetry.
 
\begin{figure*}[htb]
\includegraphics[scale=0.4]{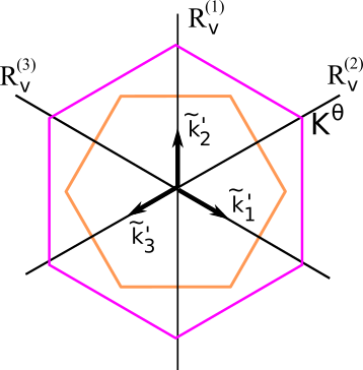}
  \caption{The BZ of graphene (purple) and of the TMDC (orange), along with  $\tilde{\mathbf{k}}'_{1,2,3}$ vectors and the reflection planes 
  $\mathcal{R}_{v}^{1,2,3}$  for $\theta=\pi/6$.
  \label{fig:30degkvect}}
\end{figure*}

 We denote the three wavevectors in the BZ of the TMDC  that satisfy the quasimomentum conservation by $\tilde{\mathbf{k}}'_{1,2,3}$, see Fig.~\ref{fig:30degkvect}.
As one can see, the $\tilde{\mathbf{k}}'_j$ vectors now lie on the $\Gamma$-$M$ lines of the BZ of the TMDC.  
 Let us consider the matrix element $\langle\Psi_{}^{(e)}(\tilde{\mathbf{k}}_1,\mathbf{r})|L_{\pm}|\Psi_{}^{(o)}(\tilde{\mathbf{k}}_1,\mathbf{r})\rangle$.
\begin{align}
 \langle\Psi_{}^{(e)}(\tilde{\mathbf{k}}'_1,\mathbf{r})|L_{\pm}|\Psi_{}^{(o)}(\tilde{\mathbf{k}}'_1,\mathbf{r})\rangle &= 
  \langle\Psi_{}^{(e)}(\tilde{\mathbf{k}}'_1,\mathbf{r})|
  (\mathcal{R}_{v}^{(3)})^{\dagger} \mathcal{R}_{v}^{(3)} L_{\pm} (\mathcal{R}_{v}^{(3)})^{\dagger} \mathcal{R}_{v}^{(3)}
  |\Psi_{}^{(o)}(\tilde{\mathbf{k}}'_1,\mathbf{r})\rangle \nonumber\\
  &= \langle\Psi_{}^{(e)}(\mathcal{R}_{v}^{(3)} \tilde{\mathbf{k}}'_1,\mathbf{r})|
  \mathcal{R}_{v}^{(3)} L_{\pm} (\mathcal{R}_{v}^{(3)})^{\dagger}
  |\Psi_{}^{(o)}(\mathcal{R}_{v}^{(3)} \tilde{\mathbf{k}}'_1,\mathbf{r})\rangle\nonumber\\
  &= \langle\Psi_{}^{(e)}(\tilde{\mathbf{k}}'_1,\mathbf{r})|
  \mathcal{R}_{v}^{(3)} L_{\pm} (\mathcal{R}_{v}^{(3)})^{\dagger}
  |\Psi_{}^{(o)}(\tilde{\mathbf{k}}'_1,\mathbf{r})\rangle\nonumber\\
  &= e^{\pm i \frac{2\pi}{3}} \langle\Psi_{}^{(e)}(\tilde{\mathbf{k}}'_1,\mathbf{r})| L_{\mp} |\Psi_{}^{(o)}(\tilde{\mathbf{k}}'_1,\mathbf{r})\rangle,
 \label{eq:SOC-offdiag-pi6}
\end{align}
where we made use of 
$\mathcal{R}_{v}^{(3)} L_{\pm} (\mathcal{R}_{v}^{(3)})^{\dagger} = e^{\pm i \frac{2\pi}{3}}L_{\mp}$ 
and that $\mathcal{R}_{v}^{(3)} \tilde{\mathbf{k}}'_{1}=\tilde{\mathbf{k}}'_{1}$.
In terms of $\alpha_{e,o}^{(\pm)}(\mathbf{k}'_1)$,  Eq.(\ref{eq:SOC-offdiag-pi6}) can be rewritten as 
$\alpha_{e,o}^{(-)}(\tilde{\mathbf{k}}'_1)=e^{-i \frac{2\pi}{3}}\alpha_{e,o}^{(+)}(\tilde{\mathbf{k}}'_1)$. Therefore, 
\begin{subequations}
\begin{align}
\alpha_{e,o}^{(x)}(\tilde{\mathbf{k}}'_1)&=\frac{1}{2}\left(\alpha_{e,o}^{(+)}(\tilde{\mathbf{k}}'_1)+\alpha_{e,o}^{(-)}(\tilde{\mathbf{k}}'_1)\right) =  
\frac{1}{2}\left(1+e^{- i \frac{2\pi}{3}}\right) \alpha_{e,o}^{(+)}(\tilde{\mathbf{k}}'_1), \\
\alpha_{e,o}^{(y)}(\tilde{\mathbf{k}}'_1)&= \frac{1}{2i}\left(\alpha_{e,o}^{(+)}(\tilde{\mathbf{k}}'_1)-\alpha_{e,o}^{(-)}(\tilde{\mathbf{k}}'_1)\right)=
\frac{1}{2 i}\left(1-e^{- i \frac{2\pi}{3}}\right) \alpha_{e,o}^{(+)}(\tilde{\mathbf{k}}'_1).
\end{align}
\label{eq:alphaxy-pi6}
\end{subequations}
It follows that $\alpha_{e,o}^{(y)}(\tilde{\mathbf{k}}'_1)=\sqrt{3} \alpha_{e,o}^{(x)}(\tilde{\mathbf{k}}'_1)$. Regarding Eq.~(\ref{eq:vartheta-tilde-def}), one finds 
%\begin{equation}
$
 {\rm Arctan}\left[\frac{{\rm Im}[e^{i\eta(\tilde{\mathbf{k}}'_1)} \alpha_{e,o}^{(x)}(\tilde{\mathbf{k}}'_1)]}{{\rm Im}[e^{i\eta(\tilde{\mathbf{k}}'_1)} 
 \alpha_{e,o}^{(y)}(\tilde{\mathbf{k}}'_1)]}\right]
= \frac{\pi}{6},
$
and from Eq.~(\ref{eq:vartheta-def})  $\vartheta_{e,o}(\tilde{\mathbf{k}}'_1)=\frac{5\pi}{6}+m'\pi=(m'+1)\pi-\pi/6$.

%%%%%%%%%%%%%%%%%%%%%%%%%%%%%%%%%%%%%%%%%%%%%%%%%%%%%%%%%%%%%%%%%%%%%%%%%%%%%%%%%%%%%%%%%%%%%%%%%%%%%%%%%%%%%%%%%%%%%%%%%%%%%%%%%%%%%%%%%%%%%%%%%%%%%%%%%%%%%%%%%%%%%%%%%%%%  

\subsubsection{$\theta=\pi/3$ twist angle}
 \label{sec:theta-pi3}

\begin{figure*}[htb]
\includegraphics[scale=0.4]{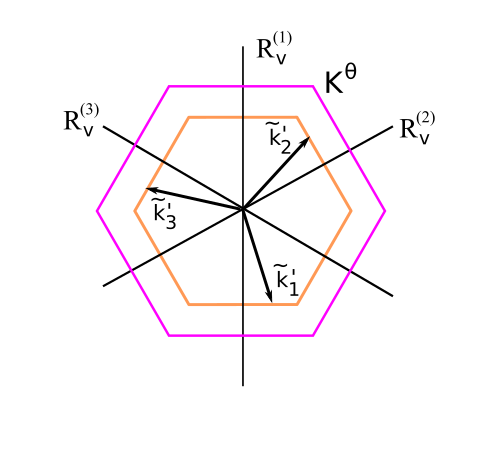}
  \caption{The BZ of graphene (purple) and of the TMDC (orange), along with  $\tilde{\mathbf{k}}'_{1,2,3}$ vectors and the reflection planes 
  $\mathcal{R}_{v}^{1,2,3}$  for $\theta=\pi/3$.
  \label{fig:60degkvect}}
\end{figure*}

  We again denote the three wavevectors in the BZ of the TMDC  that satisfy the quasimomentum conservation by 
 $\tilde{\mathbf{k}}'_{1,2,3}$, see Fig.~\ref{fig:60degkvect}. Note, that  $\tilde{\mathbf{k}}'_{1}$ is related to  $\mathbf{k}'_{1}$ in the $\theta=0$ case by  
 $\tilde{\mathbf{k}}'_{1}=\mathcal{R}_{v}^{(3)}\mathbf{k}'_{1}$ and the  $\vect{K}^{(\pi/3)}$ point of graphene is given by $-\vect{K}^{(0)}+\vect{b}_1^{0}$.
 This can be used to calculate the amplitude $t_{e(o)}(\tilde{\mathbf{k}}'_{1})$ and $\alpha_{e,o}^{(\pm)}(\tilde{\mathbf{k}}'_{1})$. Similarly to Eq.~(\ref{eq:tunneling-amp-symmetry}), 
 one may write
 \begin{align}
  t_{e(o)}(\tilde{\mathbf{k}}'_{1})e^{-i\phi_1^{X}}&=\langle\Psi^{(X)}_{gr}({\mathbf{K}}_1^{(\pi/3)},\mathbf{r})|H_{orb}|\Psi^{(e(o))}_{tmdc}(\tilde{\mathbf{k}}'_1,\mathbf{r}) \rangle
  \nonumber\\
  &=\langle\Psi^{(X)}_{gr}({-\mathbf{K}}_1^{(0)}+\vect{b}_1^{(0)},\mathbf{r})|H_{orb}|\Psi^{(e(o))}_{tmdc}(\mathcal{R}_{v}^{(3)}\mathbf{k}'_1,\mathbf{r}) \rangle\nonumber\\
  &=\langle\Psi^{(X)}_{gr}({-\mathbf{K}}_1^{(0)}+\vect{b}_1^{(0)},\mathbf{r})|H_{orb}|\Psi^{(e(o))}_{tmdc}(\mathcal{R}_{v}^{(3)}\mathbf{k}'_1,\mathbf{r}) \rangle\nonumber\\
  &=\langle\Psi^{(X)}_{gr}({-\mathbf{K}}_1^{(0)}+\vect{b}_1^{(0)},\mathbf{r})|H_{orb}|\mathcal{R}_{v}^{(3)}\Psi^{(e(o))}_{tmdc}(\mathbf{k}'_1,\mathbf{r}) \rangle\nonumber\\
   &=\langle\Psi^{(X)}_{gr}({-\mathbf{K}}_1^{(0)}+\vect{b}_1^{(0)},\mathbf{r})|\mathcal{R}_{v}^{(3)}H_{orb}|\Psi^{(e(o))}_{tmdc}(\mathbf{k}'_1,\mathbf{r}) \rangle\nonumber\\
   &=\langle\Psi^{(X)}_{gr}(\mathcal{R}_{v}^{(3)}({-\mathbf{K}}_1^{(0)}+\vect{b}_1^{(0)}),\mathbf{r})|H_{orb}|\Psi^{(e(o))}_{tmdc}(\mathbf{k}'_1,\mathbf{r}) \rangle\nonumber\\  
    &=\langle\Psi^{(X)}_{gr}({\mathbf{K}}_1^{(0)}-\vect{b}_1^{(0)}),\mathbf{r})|H_{orb}|\Psi^{(e(o))}_{tmdc}(\mathbf{k}'_1,\mathbf{r}) \rangle = 
    t_{e(o)}(\mathbf{k}'_{1})e^{-i\phi_1^{X}}e^{-i\omega_1^{X}}.
 \end{align}
This meas that $t_{e,o}(\tilde{\mathbf{k}}'_{1})=t_{e}(\tilde{\mathbf{k}}'_{1})t_{o}^{*}(\tilde{\mathbf{k}}'_{1})=t_{e}(\mathbf{k}'_{1})t_{o}^{*}(\mathbf{k}'_{1})=
t_{e,o}(\mathbf{k}'_{1})$. Regarding the matrix elements of $\hat{H}_{soc}$, one finds 
\begin{align}
 \langle\Psi_{}^{(e)}(\tilde{\mathbf{k}}'_1,\mathbf{r})|L_{\pm}|\Psi_{}^{(o)}(\tilde{\mathbf{k}}'_1,\mathbf{r})\rangle &= 
  \langle\Psi_{}^{(e)}(\mathcal{R}_v^{(3)}\mathbf{k}'_1,\mathbf{r})|L_{\pm}|\Psi_{}^{(o)}(\mathcal{R}_v^{(3)}\mathbf{k}'_1,\mathbf{r})\rangle \nonumber\\ 
  &=\langle\Psi_{}^{(e)}(\mathbf{k}'_1,\mathbf{r})|[\mathcal{R}_v^{(3)}]^{\dagger} L_{\pm}\mathcal{R}_v^{(3)}|\Psi_{}^{(o)}(\mathbf{k}'_1,\mathbf{r})\rangle\nonumber\\ 
  &= e^{\pm i \frac{2\pi}{3}}\langle\Psi_{}^{(e)}(\mathbf{k}'_1,\mathbf{r})|L_{\mp}|\Psi_{}^{(o)}(\mathbf{k}'_1,\mathbf{r})\rangle\rangle%\nonumber\\
 \label{eq:SOC-offdiag-pi3}
\end{align}
Here we have used $[\mathcal{R}_v^{(3)}]^{\dagger} L_{\pm}\mathcal{R}_v^{(3)}=e^{\pm i \frac{2\pi}{3}} L_{\mp}$. 
Performing analogous calculations as in Eqs.~(\ref{eq:alphaxy-0pi}), one finds 
\begin{equation}
 t_{e,o}(\tilde{\vect{k}}'_1)\left(
 \begin{array}{c}
  \alpha_{e,o}^{(x)}(\tilde{\mathbf{k}}'_1)  \\
  \alpha_{e,o}^{(y)}(\tilde{\mathbf{k}}'_1)  
 \end{array}
 \right) = \frac{1}{2}
 \left(
 \begin{array}{cc}
  -1 &   \sqrt{3} \\
  \sqrt{3} & 1
 \end{array}
 \right)
  \left(
  \begin{array}{c}
     \alpha_{e,o}^{(x)}(\mathbf{k}'_1) \\
     \alpha_{e,o}^{(y)}(\mathbf{k}'_1)
  \end{array}
   \right) t_{e,o}(\vect{k}'_1).
  \label{eq:alphaxy-pi3-matrix}
\end{equation}
One can now easily calculate ${\vartheta}_{e,o}(\tilde{\vect{k}}'_1)$. Using  Eq.(\ref{eq:alphaxy-pi3-matrix})
\begin{equation}
 {\rm Im}[t_{e,o}(\tilde{\vect{k}}'_1)\alpha_{e,o}^{(x)}(\tilde{\mathbf{k}}'_1)]=-\frac{1}{2} {\rm Im}[t_{e,o}(\vect{k}'_1)\alpha_{e,o}^{(x)}(\mathbf{k}'_1)]
 +\frac{\sqrt{3}}{2}{\rm Im}[t_{e,o}(\vect{k}'_1)\alpha_{e,o}^{(y)}(\mathbf{k}'_1)].
\end{equation}
However, in Sec.~\ref{sec:theta0} we found that 
$
{\rm Im}\left[t_{e,o}(\vect{k}'_1) \alpha_{e,o}^{(x)}(\mathbf{k}'_1)\right]=
\sqrt{3}\,{\rm Im}\left[t_{e,o}(\vect{k}'_1) \alpha_{e,o}^{(y)}(\mathbf{k}'_1)\right]
$, which means that ${\rm Im}[t_{e,o}(\tilde{\vect{k}}'_1)\alpha_{e,o}^{(x)}(\tilde{\mathbf{k}}'_1)]=0$ and from Eq.(\ref{eq:vartheta-def}) 
${\vartheta}_{e,o}(\tilde{\vect{k}}'_1)=2\pi/3+m''\pi=(m''+1)\pi-\pi/3$.

%%%%%%%%%%%%%%%%%%%%%%%%%%%%%%%%%%%%%%%%%%%%%%%%%%%%%%%%%%%%%%%%%%%%%%%%%%%%%%%%%%%%%%%%%%%%%%%%%%%%%%%%%%%%%%%%%%%%%%%%%%%%%%%%%%%%%%%%%%%%%%%%%%%%%%%%%%%%%%%%%%%%%%%%%%%% 

\subsection{Discussion of the results of Sec.~\ref{sec:vartheta-C3v}}
 \label{sec:theta-discuss}

Similar calculations to those in Sec.~\ref{sec:theta0}-\ref{sec:theta-pi3} can be performed for all interlayer twist angles $\theta_l=l\pi/6$, $l=0,1,2,\dots$.
We now summarize the most important  findings.
\begin{itemize}
\item One finds that  $\vartheta_{e,o}(\theta_l)$ can  be expressed as  $\vartheta_{e,o}(\theta_l)=m(\theta_l)\pi-\theta_l$, where $m(\theta_l)$ is an integer.
%%%%%%%%%%%%%
 \item The results of Sec.~\ref{sec:theta0}-\ref{sec:theta-pi3} were obtained for an arbitrary pair of $e$ and $o$ bands. This means, that using 
  Eq.~(\ref{eq:lambdaR-bar}) one finds $\vartheta_{R}(\theta_l)=n(\theta_l)\pi-\theta_l$, where $n(\theta_l)$ is an integer.
%%%%%%%%%%%%%%  
 \item According to  Eq.~(\ref{eq:H_R-transformed}),   the non-zero matrix elements of $H_R^{gr}$ are 
 $\propto\lambda_R^{}(\theta)e^{\pm i(\theta+\vartheta_{R})}$. 
       For $\theta=\theta_l$ one finds that these matrix elements are %$\lambda_R^{}(\theta_l) e^{\pm i \vartheta_{tot}(0)}=\lambda_R(\theta_l) (-1)^{(m+1)}$. 
       $\lambda_R(\theta_l) (-1)^{n(\theta_l)}$.
       This means that for interlayer twist angles  where the $C_{3v}$ symmetry of the stack is restored, the Hamiltonian of the induced Rashba SOC simplifies to 
       \begin{equation}
        %$
        H_R^{gr}=(-1)^{[n(\theta_l)+1]}\frac{{\lambda}_R(\theta_l)}{2}\left(\tau_z \sigma_x s_y - \sigma_y s_x\right),
        %$.
       \end{equation}
     which is, apart from the sign  $(-1)^{[n(\theta_l)+1]}$, the well-known result of Refs.~\cite{kane_quantum-spin-Hall_2005,min_intrinsic_2006}. This shows our results are in agreement with general expectations based on the symmetry of the system. 
  \item For twist angles $\theta\in(\theta_l,\theta_{l+1})$, i.e., when the stack has only $C_3$ symmetry,  $\vartheta_{R}$ is a continuous function of 
       $\theta$ (through the wavenumbers $\tilde{\vect{k}}'_j$). Therefore the matrix elements $\lambda_R^{}(\theta)e^{\pm i(\theta+\vartheta_{R})}$ 
       of $H_R^{gr}$ are complex numbers.
%%%%%%%%% 
 \item One would expect that ${\lambda}_R(\theta)={\lambda}_R(\theta+2\pi/3)$. As shown in Figs.~\ref{fig:MoS2-SOC-comp}(b) and \ref{fig:WSe2-SOC}(b) below, 
       we find that ${\lambda}_R(\theta)={\lambda}_R(\theta+\pi/3)$.
 Note, that Eq.(\ref{eq:Rashba-gen-appendix}) gives the  lowest order non-vanishing contribution. We expect that higher order contributions in the perturbation series, 
 albeit small in magnitude, would lead to $2\pi/3$  periodicity. 
 %%%%%%%%% 
 Note, that $\vartheta_R(\theta)\neq\vartheta_R(\theta+\pi/3)\,\,{\rm mod}\,\,2\pi$. This can be seen, e.g., comparing the results of Sec.~\ref{sec:theta0} and 
 Sec.~\ref{sec:theta-pi3}. However, the results of our  calculations show that  $\vartheta_R(\theta)+\theta$, which gives the phase of the induced Rashba SOC, does change by multiples 
 of $2\pi$ upon interlayer rotation by $\theta=\pi/3$. 
\end{itemize}

%%%%%%%%%%%%%%%%%%%%%%%%%%%%%%%%%%%%%%%%%%%%%%%%%%%%%%%%%%%%%%%%%%%%%%%%%%%%%%%%%%%%%%%%%%%%%%%%%%%%%%%%%%%%%%%%%%%%%%%%%%%%%%%%%%%%%%%%%%%%%%%%%%%%%%%%%%%%%%%%%%%%%%%%%%%%  

{\section{Details of the numerical calculations for $\vartheta_{R}$,  $\lambda_R^{}$ and $\lambda_{vZ}^{}$}
\label{sec:corrections}}

In this section  we discuss  questions related to the choice of material and other parameters in our numerical calculations and compare some of our 
results to corresponding results in Ref.~\cite{david_induced_2019}.

\vspace{0.3cm}

{\subsection{Interlayer tunneling parameters}
\label{sec:tunneling-coeffs}}

In Ref.~\cite{david_induced_2019} it was argued that for the description of the tunneling between the monolayer graphene (MLG) sheet and the 
MoS$_2$ layer %, i.e., in the calculation of the amplitude $t_b$,  
it is sufficient to consider only the $p$ orbitals of the chalcogen atoms and one can neglect  
the  $d$ orbitals of the metal atoms. Namely, the  the  $d$ orbitals are farther away from the carbon atoms of MLG and therefore their overlap
with the carbon $p_z$ orbitals should be small.   %To check the results of this approach,  
We have extended the  calculations of   Ref.~\cite{david_induced_2019}  by  taking into account  contributions from tunneling processes involving the $d$ orbitals 
of the metal atoms  as well. The results shown in the main text have been obtained in this way. 

As already mentioned in Sec.~\ref{sec:intro-to-vartheta}, the tunneling matrix $(T({\tau\vect{k}'_j}))_{b}$ can be written as 
\begin{equation*}
 \left(T(\tau \bm{k}'_j)\right)_b= t_b(\tau\vect{k}'_j) e^{i\tau \bm{G}^\theta_j \cdot \bm{r}_0} 
\left(\begin{array}{c} e^{i\tau \phi_j^{A}} \\ e^{i\tau \phi_j^{B}}\end{array}\right).
\end{equation*}
When only the  $p$ orbitals are  considered in the interlayer tunneling, then 
the interlayer tunneling amplitude $t_b^{p}(\tau\mathbf{K}^\theta,\tau\vect{k}'_j)\equiv t_b^{p}(\tau\vect{k}'_j)$ 
between  the $\mathbf{K}^{\theta}$ point of graphene and   
electronic states at wavenumber $\mathbf{k}'_j$ in band $b$ of the TMDC is given by%\cite{david_induced_2019}
\begin{equation}
  t_b^{p}(\tau\vect{k}'_j) =
      i \tau [c_{bx} (\tau\mathbf{k}'_j) \cos\theta + c_{by} (\tau\mathbf{k}'_j) \sin\theta] \; t_{\parallel}^{p} + c_{bz} (\tau\mathbf{k}'_j) \; t_{\perp}^{p}.      
\label{eq:p-orb-tunneling}      
\end{equation}  
Here $c_{bx} (\tau\mathbf{k}'_j)$, $c_{by} (\tau\mathbf{k}'_j)$ and $c_{bz} (\tau\mathbf{k}'_j)$ are the complex amplitudes of the $p_x$, $p_y$ and $p_z$ orbitals of 
the chalcogen atoms  at the BZ point $\mathbf{k}'_j$ in band $b$ of the TMDC. As it was shown in Ref.~\cite{david_induced_2019}, 
$t_b^{p}(\tau\vect{k}'_{2,3})=t_b^{p}(\tau\vect{k}'_{1})$. 
According to  Eq.~(\ref{eq:p-orb-tunneling}), 
two independent tunneling coefficients are needed: $t_{\parallel}^{p}$ and $t_{\perp}^{p}$,  and Ref.~\cite{david_induced_2019} obtained them from a fitting procedure using  previous results~\cite{Gmitra_TMDC-proximity_2016} of  DFT band structures calculations  for MLG/TMDC heterostructures. 
The values of these parameters were estimated to be  $t_{\parallel}^{p} \approx t_{\perp}^{p} \approx 100\,$meV. 

 Considering now the $d$ orbitals of the TMDC metal atoms one can derive a similar expression 
 for the  tunneling amplitude $t_b^d(\tau\vect{k}'_j)$ between MLG and  the $d$ orbitals:
\begin{eqnarray}
t_b^d(\tau\vect{k}'_{2,3})=
t_b^d(\tau\vect{k}'_1) &=& \left(c_{b,x^2-y^2}(\tau{\bm k'_1})\cos(2\theta)+c_{b,xy}(\tau{\bm k'_1})\sin(2\theta)\right)t^d_{\square}+\nonumber\\
& & i\tau\left(c_{b,xz}(\tau{\bm k'_1})\cos(\theta)+c_{b,yz}(\tau{\bm k'_1})\sin(\theta)\right)t^d_{||}+c_{b,3z^2-r^2}(\tau{\bm k'_1})t^d_\perp.
\label{eq:d-orb-tunneling}
\end{eqnarray}
Here $c_{b, d}(\tau{\bm k'_1})$  ($d=\{x^2-y^2, xy, xz, yz,3z^2-r^2\}$) are the complex amplitudes of the $d$ orbitals of the TMDC metal atoms  at the BZ point 
$\mathbf{k}'_1$ in band $b$.  The derivation of Eq(\ref{eq:d-orb-tunneling}) 
involves the same steps as that of Eq.~(\ref{eq:p-orb-tunneling}) and therefore we do not give further details here. 
As one can see,  $t_b^d(\tau\vect{k}'_1)$ involves three 
more overlap parameters: $t^d_{\square}$, $t^d_{||}$, $t^d_\perp$. Therefore  there are altogether five $t^{p, d}$  parameters  that  describe the 
overlap between graphene's  $p_z$ orbitals and the $p$, $d$ orbitals of the TMDC layer. The approach used in Ref.~\cite{david_induced_2019} 
to estimate $t_{\parallel}^{p}$ and $t_{\perp}^{p}$ is not  applicable to estimate all five  $t^{p, d}$  parameters. 

Therefore we  used   the Slater--Koster method to re-calculate the tunneling coefficients both for the chalcogen $p$ orbitals and the $d$ orbitals 
of the metal atoms. The steps needed to calculate  $t_{\parallel}^{p}$, $t_{\perp}^{p}$, $t^d_{\square}$, $t^d_{||}$, and $t^d_\perp$ in terms of the set of parameters
of the Slater--Koster method are explained in Refs.~\cite{david_induced_2019, li_twist-angle_2019}. %, therefore we do not repeat them here. 
We only note that the parametrization of the transfer integrals is based on the Harrison's model\cite{harrison_elementary_1999}  and the numerical values of the 
necessary parameters, with the exception of $\eta_{\rm C-Se}$,  are listed in Ref.\cite{li_twist-angle_2019}, see Eqs(4), (5) therein. All our calculations for MLG/MoS$_2$ 
are based on  the  parameters  given in Ref.\cite{li_twist-angle_2019}. 

However, for the calculations for MLG/WSe$_2$ we also needed the Harrison's model parameter $\eta_{\rm C-Se}$.
We have fixed its value in the following  way. We took the values $\lambda_{vZ}$ and  $\lambda_R$ of the induced valley Zeeman  and Rashba SOC 
from Ref.~\cite{Gmitra_TMDC-proximity_2016}, where 
they were obtained by fitting band structure calculations of graphene/WSe$_2$ heterostructures at $\theta=0$ twist angle. 
Using the  methodology of Ref.~\cite{david_induced_2019} to calculate $\lambda_{vZ}$ and  $\lambda_R$,  we adjusted the value of $\eta_{\rm C-Se}$ such that 
we obtain a reasonable agreement with Ref.~\cite{Gmitra_TMDC-proximity_2016}, for $\theta=0$. 
We found that $\eta_{\rm C-Se}=0.748$ minimizes the deviation to $\approx  \pm 20\%$  between our results for $\lambda_{vZ}$, $\lambda_R$  and those of  Ref.~\cite{Gmitra_TMDC-proximity_2016}. 

\begin{table}[ht]
 \begin{tabular}{|c|c|c|c|c|c|}
 \hline
    &  $t_{\parallel}^{p}$ [meV] & $t_{\perp}^{p}$  [meV]&  $t^d_{\square}$  [meV]& $t^d_{||}$ [meV]& $t^d_\perp$ [meV]\\    
  \hline
    MoS$_2$ & $-143.6$   &  $144.7$ &  $9.2$ & $17.6$ & $-14.9$\\  
    WSe$_2$ &  $-76.2$  & $76.2$ &  $2.2$ & $4.1$ & $-3.5$\\
    \hline
 \end{tabular}
\caption{Tunneling coefficients}
\label{tbl:t-coeffs}
\end{table}
To summarize, we used  tunneling coefficients shown in Table~\ref{tbl:t-coeffs}.
Regarding the $t^{p}$ coefficients for MoS$_2$, they are indeed not very different from those estimated in Ref.~\cite{david_induced_2019} ($\approx 100\,$meV) using 
a different method, but the sign of  $t_{||}^{p}$ is opposite to the one given in Ref.~\cite{david_induced_2019}. 
This can be traced back to the fact that the TB model of Ref.\cite{fang_ab_2015}  uses a different orientation of the TMDC lattice with respect to Ref.~\cite{david_induced_2019}.
This gave rise to an overlooked $\vect{k} \to -\vect{k}$ transformation in reciprocal space, which resulted in a sign change in the estimation of the tunneling parameter.
One can also see that the  values of the  $t^d$ coefficients are indeed much smaller than $t_{\parallel}^{p}$ and $t_{\perp}^{p}$. 
On the other hand, the complex amplitude factors $c_{b,x^2-y^2}$, $c_{b,xy}$, $c_{b,xz}$, $c_{b,yz}$, 
and $c_{b,3z^2-r^2}$ turn out to be significantly larger than $c_{bx}$, $c_{by}$ and $c_{bz}$ in large regions of the BZ. Therefore 
$t_b^d(\tau\mathbf{k}'_1)$ is found to be small, but  non-negligible  with respect to $ t_b^{p} (\tau\mathbf{k}'_1)$.

%%%%%%%%%%%%%%%%%%%%%%%%%%%%%%%%%%%%%%%%%%%%%%%%%%%%%%%%%%%%%%%%%%%%%%%%%%%%%%%%%%%%%%%%%%%%%%%%%%%%%%%%%%%%%%%%%%%%%%%%%%%%%%%%%%%%%%%%%%%%%%%%%%%%%%
%\vspace{0.3cm}

\subsection{Further parameters}
\label{sec:further-params}

In addition to $t_b(\tau\mathbf{k}'_j)$, there are a couple of further parameters that enter the numerical calculations. We will briefly discuss them here. 
\begin{itemize}
\item The value of the band gap $E_g$ of the TMDC. 

 As briefly discussed in the main text, one can take its value either from experiments (when available) 
 or from previous theoretical works. Regarding the latter,  since we use the TMDC TB model  of Ref.~\cite{fang_ab_2015} to calculate the band structure, 
 the spin-orbit coupling Hamiltonian matrix elements and the  interlayer tunneling amplitude, we also used the   $E_g$ of this model. 
 (We implemented the TB model in the Kwant code\cite{groth_kwant_2014}). Although  the model of Ref.~\cite{fang_ab_2015} itself is based on DFT calculations, 
 the value of $E_g$ is different from what one can extract from the calculations of Ref.~\cite{Gmitra_TMDC-proximity_2016} that were performed for 
 MLG/TMDC supercells at  $\vartheta=0$. For example, in the case of MoS$_2$ the  model of Ref.~\cite{fang_ab_2015} has a band gap that is $17\%$ larger  than 
 the corresponding $E_g$  given in Ref.~\cite{Gmitra_TMDC-proximity_2016}. See Table~\ref{tbl:further-params} for the $E_g$ values used in this work. 

\item The position of the Dirac point of graphene within the the band gap of the TMDC. 

We describe the energy of the Dirac point of graphene in
the band gap of the TMDC by a number $f_G \in [0, 1]$. Its value is a linear function of the position of the Dirac point in the TMDC band gap. When $f_G = 0$, 
the Dirac point is aligned with the TMDC valence band edge, for $f_G = 1$ the Dirac point has the same energy as the TMDC conduction band edge. 
When available, the value of $f_G$ can also be taken from experiments\cite{pierucci_graphene-MoS2,nakamura_graphene-WSe2}. 
See Table~\ref{tbl:further-params} for the $f_g$ values used in this work. 

\item The number of bands in the model for the TMDC layer. 

The explicit expression to calculate  $\lambda_{vZ}^{}$ involve a sum over contributions of individual bands. Similarly, in order to calculate
$\vartheta_{R}$ and  $\lambda_R^{}$, one needs to sum over  pairs of even (e) and odd (o) bands. 
Ref.~\cite{david_induced_2019}  used the approximation that  
the conduction band (CB) and the valence band (VB) were taken into account for $\lambda_{vZ}$, and three pairs of e-o bands for $\lambda_R$. 
As it will be shown below,  this  already  leads to qualitatively good results in most cases, see Sec.~\ref{sec:bls-SOC-numerics} below.
An exception is the value of $\lambda_R$ for $\theta\approx \pi/6$, where, due  to band crossings and  near-degeneracies of certain $e$ bands, 
more pairs of $e$ and $o$ bands need to be taken into account.   
The TB model that we use involves $11$ bands of the TMDC layer.  Unless otherwise indicated,  we  use the contributions of all $11$ bands 
 to calculate $\lambda_{vZ}$,  and $30$ pairs of $e$ and $o$ bands for the $\vartheta_{R}$ and $\lambda_R$ calculations.  
\begin{table}[ht]
 \begin{tabular}{|c|c|c|c|c|}
 \hline
      &  $E_g$ (DFT)   &  $E_g$ (exp) &  $f_G$ (DFT) &  $f_G$ (exp) \\
 \hline
    MoS$_2$ & $1.807\,$eV\cite{fang_ab_2015}   &  $2.0\,$eV\cite{pierucci_graphene-MoS2} &  $0.974\,$ \cite{Gmitra_TMDC-proximity_2016}& $0.55$\cite{pierucci_graphene-MoS2} \\  
    WSe$_2$ & $1.638\,$eV\cite{fang_ab_2015}   &  $1.95\,$eV\cite{nakamura_graphene-WSe2} &  $0.161$\cite{Gmitra_TMDC-proximity_2016} & $0.426$\cite{nakamura_graphene-WSe2} \\
 \hline
 \end{tabular}
\caption{Band gap and MLG Dirac point energy position parameters from  DFT calculation and experiments.}
\label{tbl:further-params}
\end{table}

\item The TB model of the TMDC

We also note that the TB model of the TMDC appears to be less accurate for the bands above the band gap\cite{fang_ab_2015}. 
This can add further uncertainty to the results, especially  in the case graphene/MoS$_2$, where  $f_G=0.974$, i.e., 
the relative contributions of $e$-$o$ band pairs above $E_g$ is larger than the contribution coming from the valence bands. 
\end{itemize}

These details, in addition to those already discussed in Sec.~\ref{sec:tunneling-coeffs}, also need to be considered  when comparing our results  to  
related previous works\cite{Gmitra_TMDC-proximity_2016,li_twist-angle_2019,david_induced_2019,Pezo2020twist}.

%%%%%%%%%%%%%%%%%%%%%%%%%%%%%%%%%%%%%%%%%%%%%%%%%%%%%%%%%%%%%%%%%%%%%%%%%%%%%%%%%%%%%%%%%%%%%%%%%%%%%%%%%%%%%%%%%%%%%%%%%%%%%%%%%%%%%%%%%%%%%%%%%%%%%%%%%%

\subsection{Numerical calculations of the induced SOC in MLG/TMDC twisted bilayers}
\label{sec:bls-SOC-numerics}

We now show our results  for the twist angle dependence of  $\lambda_{vZ}$ and $\lambda_R$ for graphene/MoS$_2$ and graphene/WSe$_2$. In the former case we also 
compare our results to those of Ref.~\cite{david_induced_2019}. For the numerical calculations of $\lambda_R$ we evaluated Eqs.~(\ref{eq:lambdaR_eo}) and (\ref{eq:lambdaR-bar}).
To calculate $\lambda_{vZ}$ we used the result of Ref.~\cite{david_induced_2019}:
\begin{equation}
\lambda_{\rm VZ} = 3\sum_b \frac{ \left| t_b(\bm{k}'_1) \right|^2 \Delta_{0,b}(\bm{k}'_1)}{E_b^2(\bm{k}'_1) - \Delta_{0,b}^2(\bm{k}'_1)},
\label{eq:lambdaVZ}
\end{equation}
where $t_b(\bm{k}'_1)$ is defined in Eq.~(\ref{eq:tunneling-amplitude-def}) and the summation runs over all bands $b$ of the tight-binding model.

%%%%%%%%%%%%%%%%%%%%%%%%%%%%%%%%%%%%%%%%%%%%%%%%%%%%%%%%%%%%%%%%%%%%%%%%%%%%%%%%%%%%%%%%%%%%%%%%%%%%%%%%%%%%%%%%%%%%%%%%%%%%%%%%%%%%%%%%%%%%%%%%%%%%%%%%%%%%

\subsubsection{Comparison to the results of Ref.~\cite{david_induced_2019}}

\begin{figure}
\begin{center}
	\igr[scale=0.23]{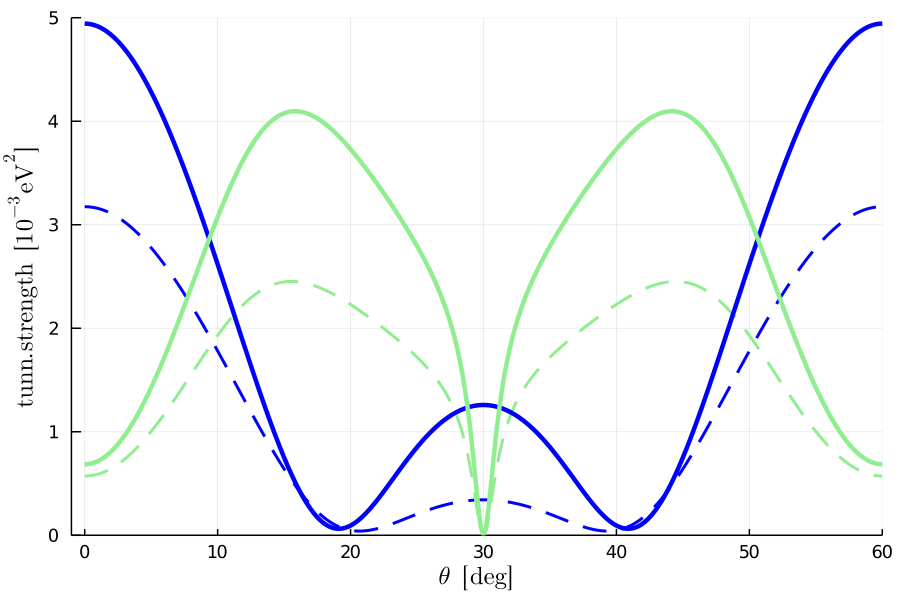}
%	\igr[scale=0.23]{3c.png}
	\caption{Tunneling strength $|t_c|^2$ to the conduction band  (light green) and to the valence band $|t_{v}|^2$ (blue) of monolayer MoS$_2$. 
	The solid curves are the results of this work, the dashed curves are reproduced from Ref.~\cite{david_induced_2019}. }
\label{fig:tunneling-strenght-comp}
\end{center}
\end{figure}

First, we briefly discuss how the tunneling  parameters  shown in Table~\ref{tbl:t-coeffs}  modify the results for tunneling amplitude $t_b$  compared to the 
corresponding results in Ref.~\cite{david_induced_2019}. 
In Fig.~\ref{fig:tunneling-strenght-comp} we  plot the tunneling strength  $|t_c|^2$ and $|t_{v}|^2$ to the CB and VB of MoS$_2$ 
as a function of the twist angle $\theta$ using the parameters from Table~\ref{tbl:t-coeffs}  (solid lines). We also show  the corresponding 
results of  Ref.~\cite{david_induced_2019}, where only  $t_{\parallel}^{p} \approx t_{\perp}^{p}$ were non-zero. 
As one can see in Fig.~\ref{fig:tunneling-strenght-comp}, the present calculations tend to yield larger tunneling strengths. 
However, the twist angle dependence of  $|t_c|^2$ and $|t_{v}|^2$ remain qualitatively the  same.

\begin{figure}
\begin{center}
     \igr[scale=0.4]{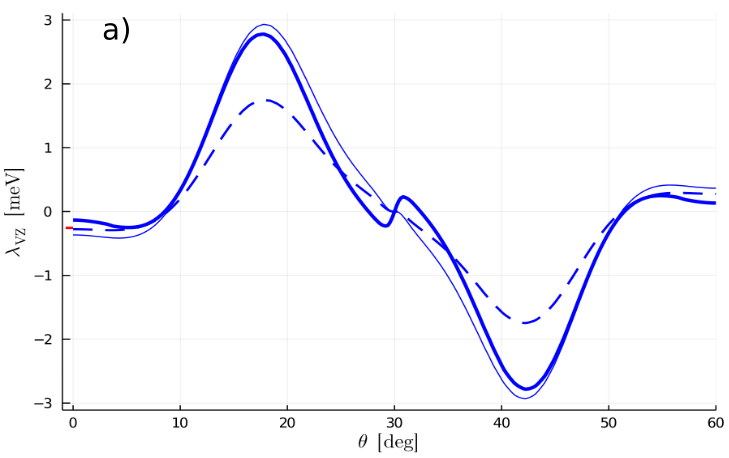}
     \igr[scale=0.4]{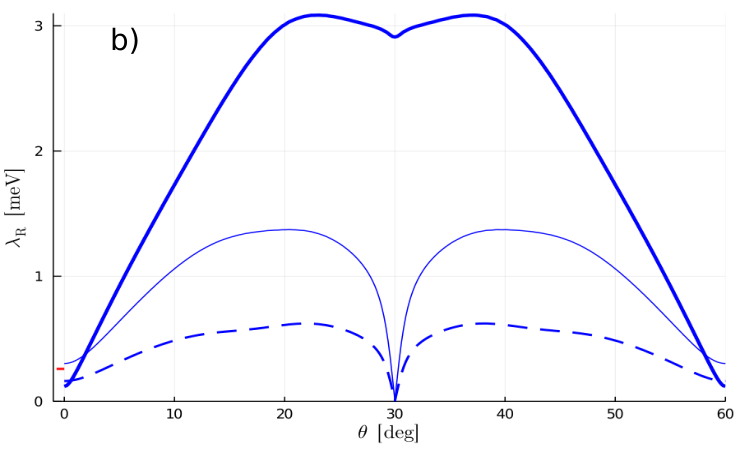}
	\caption{a) $\lambda_{vZ}$ and b) $\lambda_R$ for MLG/MoS$_2$ as a function of interlayer twist angle $\theta$ using the DFT parameters given in Table~\ref{tbl:further-params}.
	 Different lines correspond to different tunneling amplitudes and number of bands in the calculations, see text for details. 
	 The red marks at $\theta = 0$ indicate the results of Ref.~\cite{Gmitra_TMDC-proximity_2016}. }
\label{fig:MoS2-SOC-comp}
\end{center}
\end{figure}

In  Fig.~\ref{fig:MoS2-SOC-comp}(a)  we show calculations for $\lambda_{vZ}$ vs $\theta$ for graphene/MoS$_2$ using the DFT parameters for $E_g$  and $f_G$, 
see Table~\ref{tbl:further-params}.
The dashed curve in Fig.~\ref{fig:MoS2-SOC-comp}(a)  again corresponds to the result of Ref.~\cite{david_induced_2019}, which uses the contributions of the CB and the VB only. 
The thin solid line is  obtained using the  tunneling amplitudes from Table~\ref{tbl:t-coeffs} and only tunneling to the  VB and the CB is taken into account.
One can see a pronounced increase of $\lambda_{\rm VZ}$ around $\theta=18^{\circ}$ with respect to Ref.~\cite{david_induced_2019}, which is mainly due to the 
larger tunneling strengths, as shown in Fig.~\ref{fig:tunneling-strenght-comp}. Nevertheless,  the curves remain qualitatively very similar. 
The thick solid line shows the result that we obtain by using the  tunneling amplitudes from Table~\ref{tbl:t-coeffs} and 
taking into account  tunneling to all $11$  bands of the TB model. It is again qualitatively similar to the other 
two results, the most important change is 
the two new zero crossings close to $\theta=30^\circ$. 

In  Fig.~\ref{fig:MoS2-SOC-comp}(b)  we show similar calculations for $\lambda_{R}$ vs $\theta$.  
The larger tunneling strength $t_b$ used in  this work leads again to an enhancement of  $\lambda_{R}$ (thin solid line) compared  to Ref.~\cite{david_induced_2019} (dashed line), 
but the $\theta$ dependence of the two results otherwise agree, including the deep minimum at $\theta=30^{\circ}$. However, if all possible $e$-$o$ pairs of bands are included 
in the calculation (thick solid line), then  this  minimum becomes a small dip and the value of $\lambda_{R}$ is significantly enhanced. 
As already mentioned, this is  because of  band crossings and  near-degeneracies of certain $e$ bands which means that more pairs of $e$ and $o$ bands need to be taken 
into account than it was done in Ref.~\cite{david_induced_2019}.

For comparison, in Fig.~\ref{fig:MoS2-SOC-comp} we also indicate by  red marks the results of the DFT calculations of Ref.~\cite{Gmitra_TMDC-proximity_2016} for $\theta=0$.
%$\lambda_{\rm vZ} \approx -0.255\,$meV. $2\lambda_{\rm R}=0.26\,$meV,
As one can see, all our curves  take values  close to these   reference values. % at $\theta=0^\circ$. 
As we explained in Sec.~\ref{sec:further-params}, the exact results depend on a number of details, including potentially fine-tuned Slater--Koster parameters. 
This leaves us with much freedom to adjust our parameters and fit previously published results, but this goes beyond the scope of this paper.

We found a qualitatively similar $\lambda_{vZ}$ and $\lambda_{R}$ vs $\theta$ dependence using the  experimental  parameters of $E_g$  and $f_G$ from 
Table~\ref{tbl:further-params}, but the maximum values of the induced SOC are significantly  smaller.

%%%%%%%%%%%%%%%%%%%%%%%%%%%%%%%%%%%%%%%%%%%%%%%%%%%%%%%%%%%%%%%%%%%%%%%%%%%%%%%%%%%%%%%%%%%%%%%%%%%%%%%%%%%%%%%%%%%%%%%%%%%%%%%%%%%%%%%%%%%%%%%%%%%%%%%%%%%%%%%%%%%%%%%%%%%%%%%%%%%%%%

\vspace{0.3cm}

\subsubsection{Calculations for MLG/WSe$_2$}

\begin{figure}
	\includegraphics[scale=0.4]{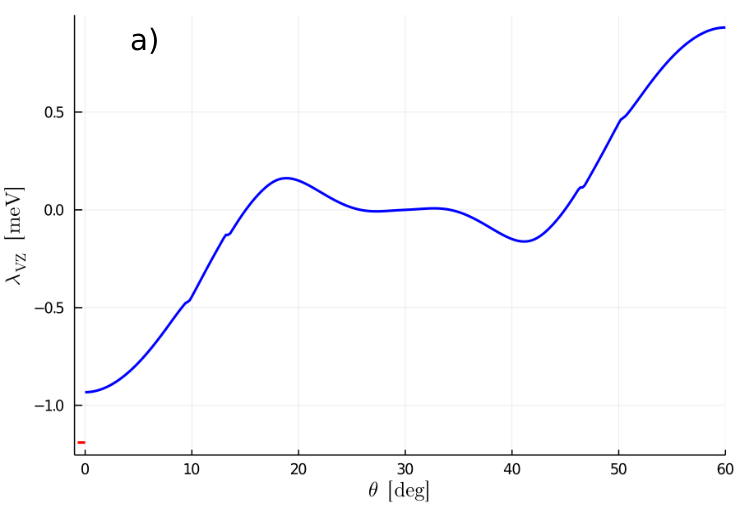}
	\includegraphics[scale=0.4]{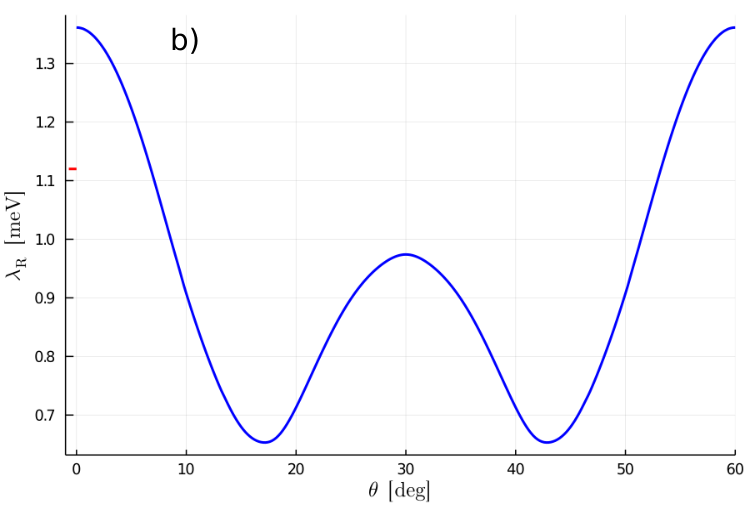}
	\caption{a) $\lambda_{vZ}$ and b) $\lambda_R$ for MLG/WSe$_2$ as a function of interlayer twist angle $\theta$ 
	using the DFT parameters given in Table~\ref{tbl:further-params}.
	 The red marks at $\theta = 0$ indicate the results of Ref.~\cite{Gmitra_TMDC-proximity_2016}.}
	\label{fig:WSe2-SOC}
\end{figure}

Our calculations for the twist angle dependence of $\lambda_{vZ}$ and $\lambda_R$ are shown in Fig.~\ref{fig:WSe2-SOC}. We used the tunneling parameters given in 
Tables~\ref{tbl:t-coeffs} and the DFT parameters in Table~\ref{tbl:further-params}.  Contributions from all bands or pairs of bands of the TB model are taken into account. 
Red marks  denote  the results of Ref.~\cite{Gmitra_TMDC-proximity_2016} for $\theta=0$, they differ $\approx 20\%$ from our results. 

Importantly, one can see  that both  $\lambda_{vZ}(\theta)$ and $\lambda_R(\theta)$  are different from the MLG/MoS$_2$ heterostructures  in Fig.~\ref{fig:MoS2-SOC-comp}. 
This is mainly due to the fact that the Dirac point of graphene is closer to the VB  (we used $f_G=0.161$ in these calculations versus $f_G=0.976$ in Fig.~\ref{fig:MoS2-SOC-comp}, 
i.e., here  the valence bands give  larger contributions. 
Similarly to MLG/MoS$_2$,  $\lambda_{vZ}$ and $\lambda_{R}$ vs $\theta$ dependence using the  experimental  parameters for $E_g$  and $f_G$ from 
Table~\ref{tbl:further-params} are qualitatively similar, but the maximum values of the induced SOC are significantly  smaller. 

\vspace{0.3cm}

\emph{Comparison of MLG/MoS$_2$  and MLG/WSe$_2$}

The different $\theta$ dependence of the induced SOC for MLG/MoS$_2$ (Fig.~\ref{fig:MoS2-SOC-comp}) and MLG/WSe$_2$ (Fig.~\ref{fig:WSe2-SOC}) explain the  
differences in the results for the corresponding twisted  trilayer systems shown in Fig.~3 of the main text. We point out, in particular, that $\lambda_R$ has a single maximum 
for  MLG/MoS$_2$,  while it has two maxima for  MLG/WSe$_2$. Together with the $\theta$ dependence of $\vartheta_{R}$ given in Fig.~1 of the main text, this 
can explain   the finding  that  $\lambda_R^{(tls)}$ can be enhanced for WSe$_2$/MLG/WSe$_2$, but not for MoS$_2$/MLG/MoS$_2$.

%%%%%%%%%%%%%%%%%%%%%%%%%%%%%%%%%%%%%%%%%%%%%%%%%%%%%%%%%%%%%%%%%%%%%%%%%%%%%%%%%%%%%%%%%%%%%%%%%%%%%%%%%%%%%%%%%%%%%%%%%%%%%%%%%%%%%%%%%%%%%%%%%%%%%%%%%%%%%%

\subsubsection{Comparison to the results of Refs.~\cite{Naimer2021twistangle} and \cite{Pezo2020twist}}

During the preparation of this manuscript, a publication  by Naimer \emph{et.~al.}\cite{Naimer2021twistangle} has appeared discussing the twist angle dependence
of the proximity-induced SOC in graphene/TMDC bilayers.  Their  methodology  is based on DFT calculations.  The incommensurate graphene/TMDC heterostructures  
were approximated  by periodic supercells, which required several approximations, e.g.,  straining the graphene sheet.

In agreement with our results, the calculations of Ref.~\cite{Naimer2021twistangle} 
indicate that the strength of the proximity induced SOC in graphene/TMDC heterostructures is tunable by interlayer twist. 
The twist angle dependence of $\lambda_{vZ}$ for graphene/MoS$_2$ and graphene/WSe$_2$ are qualitatively similar: in the former case there is a maximum at $\theta\approx 20^{\circ}$, in the latter case $\lambda_{vZ}$  monotonically decreases as 
$\theta$ is changed from $0^{\circ}$ to $30^{\circ}$. 
They also find that the phase factor  $\vartheta_R$ in $H_R$ is in general non-zero and  depends on the twist angle $\theta$.
Regarding the $\theta$ dependence
of $\lambda_R$, for  graphene/MoS$_2$ their result seems to be qualitatively similar to ours, while for 
graphene/WSe$_2$ the results show  differences. 
On a more quantitative level, however, our results for $\lambda_{vZ}$ and $\lambda_R$ display a much stronger enhancement as a function of $\theta$ than in Ref.~\cite{Naimer2021twistangle}.  The reasons for this discrepancy remains to be investigated in future works.

Another relevant recent work is by Pezo \emph{et. al.}\cite{Pezo2020twist}, who  performed supercell based DFT calculations for graphene/MoTe$_2$ bilayers 
for three different  interlayer twist angles ($\theta=0,\pi/12,\pi/6$). It was noted in  Ref.~\cite{Pezo2020twist} that the position of graphene's 
Dirac point is closer to the valence band of MoTe$_2$. Therefore one would expect that the $\theta$ dependence of the induced SOC should be similar 
to the one in graphene/WSe$_2$, meaning, e.g.,  that magnitude of both $\lambda_{vZ}$ and $\lambda_R$ should 
decrease for $\theta=\pi/6$ compared to their $\theta=0$ value. This seems to be in agreement with  results of Ref.~\cite{Pezo2020twist}.

Finally, we note that both  Refs.~\cite{Naimer2021twistangle} and \cite{Pezo2020twist} emphasize the effect of the strain that is always present 
in their supercell based DFT calculations. 

%%%%%%%%%%%%%%%%%%%%%%%%%%%%%%%%%%%%%%%%%%%%%%%%%%%%%%%%%%%%%%%%%%%%%%%%%%%%%%%%%%%%%%%%%%%%%%%%%%%%%%%%%%%%%%%%%%%%%%%%%%%%%%%%%%%%%%%%%%%%%%%%%%%%%%%%%%%%%%%%%%%%%%%%%%%%%%%%%%%

\section{Supermoir\'e effects}

\begin{figure}
\begin{center}
	\includegraphics[scale=0.7]{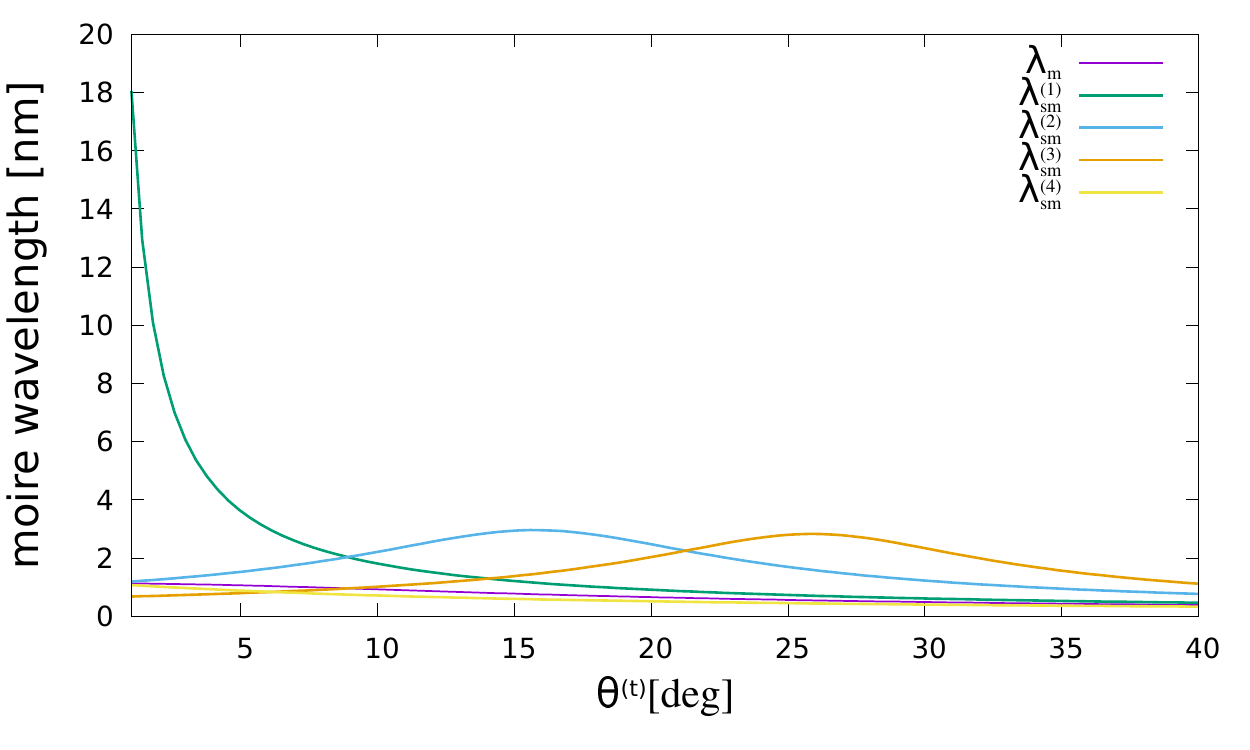}
	\caption{Moir\'e wavelength $\lambda_m$ calculated from Eq.~(\ref{eq:moire}) and supermoir\'e wavelengths calculated from Eqs.~(\ref{eq:supermoire})
	as a function of the interlayer twist angle.}
	\label{fig:moire}
\end{center}	
\end{figure}

Let us first consider the moir\'e effects in a graphene/TMDC  system. 
One can define the primitive moir\'e wavelength $\lambda_m$ for a hexagonal heterostructure\cite{Yankowitz2012} consisting of graphene and a substrate as 
\begin{equation}
 \lambda_m=\frac{q}{\sqrt{1+q^2-2 q \cos\theta}}\,a_{gr},
 \label{eq:moire}
\end{equation}
where $q=a_{s}/a_{gr}$ is the ratio of the lattice constants of the substrate ($a_{s}$) and of graphene ($a_{gr}$) and $\theta$ is the interlayer twist angle. 
In the case of graphene and e.g., monolayer MoS$_2$ this ratio is $q\approx 1.283$. 
The maximum of $ \lambda_m$  can be found for $\theta=0$ where  $\lambda_m\approx 1.1$\,nm (see Fig.\ref{fig:moire}). %, in other words, to roughly four unit cell of graphene. 
%This  can be hardly called a Moir\'e pattern in the usual sense of this  word. 
As a comparison,  the moir\'e length scale is $\lambda_m=13.9$\,nm for a graphene/hBN bilayer at $\theta=0$. %, i.e., ten times larger. 

The moir\'e  potential  due to  this  periodic perturbation leads to, e.g., gap openings at the moir\'e
Brillouin zone boundaries\cite{Yankowitz2012} 
at energies $E_M=\pm \frac{\hbar v_F G_M}{2}$, where  $G_M=\frac{4\pi}{\sqrt{3}\lambda_m}$. Using $v_F=10^6$m/s, 
for graphene/MoS$_2$ this energy scale is $E_M\approx \pm 2.2$eV for $\theta=0$, which is  far from the Dirac point. 
Indeed, in the ARPES experimental results of Ref.~\cite{pierucci_graphene-MoS2} on graphene/MoS$_2$,  
superlattice effects were observed at binding energies $E_b=-3.55$eV, from which  the authors concluded that the interlayer 
twist angle was $\theta\approx 6^{\circ}$ in their sample. 
In short,  because of the relatively large difference between the graphene and TMDC lattice constants, 
no moir\'e effects are to be expected close to the Dirac point of graphene for any  twist angle $\theta$ in a graphene/TMDC system. 

The situation might be  different in TMDC/graphene/TMDC trilayers. Namely, studying hBN/graphene/hBN trilayers, 
Refs.~\onlinecite{Peeters-supermoire,Koshino-supermoire} found that band gaps can appear in the spectrum of these heterostructures at energies closer 
to the Dirac point of graphene than what was found in hBN/graphene bilayers. This can be understood as a consequence of an interference of the  moir\'e patterns 
coming from the two hBN layers.  Ref.~\cite{Peeters-supermoire} found that if one of the interlayer twist angles, e.g., $\theta^{(b)}$   
is kept fixed at $\theta^{(b)}=0$ and $\theta^{(t)}$ is changed, then the four longest wavelength component  $\lambda_{sm}^{(i)}$ 
of such a \emph{supermoir\'e} potential in graphene are given by
\begin{subequations}
\begin{align}
 \lambda_{sm}^{(1)}&=\frac{q}{\sqrt{2-2\cos\theta^{(t)}}}\, a_{gr} \label{eq:supermoire-1},\\
  \lambda_{sm}^{(2)}&=\frac{q}{\sqrt{(2-\delta)(1-\cos\theta^{(t)})+\delta^2-\sqrt{3}\delta\sin\theta^{(t)}}}\, a_{gr},\\ 
  \lambda_{sm}^{(3)}&=\frac{q}{\sqrt{2-2\cos\theta^{(t)}+3 \delta^2-2 \sqrt{3}\delta\sin\theta^{(t)}}}\, a_{gr},\\
  \lambda_{sm}^{(4)}&=\frac{q}{\sqrt{(2-\delta)(1-\cos\theta^{(t)})+\delta^2+\sqrt{3}\delta\sin\theta^{(t)}}}\, a_{gr},
  \label{eq:supermoire-4}
\end{align}
\label{eq:supermoire}
\end{subequations}
where we used the notation $q-1=\delta$. Eqs.~(\ref{eq:supermoire})   are valid also for TMDC/graphene/TMDC 
trilayers because the constituent layers have hexagonal lattices.  We plot $\lambda_{m}^{}$  as well as $\lambda_{sm}^{(i)}$ as a function of $\theta^{(t)}$ 
using the lattice constants of graphene and MoS$_2$ in Fig.~\ref{fig:moire}.
As one can see,   $\lambda_{sm}^{(2)}$, $\lambda_{sm}^{(3)}$, $\lambda_{sm}^{(4)}$  are always quite short. However, $\lambda_{sm}^{(1)}$ does increase when 
the two TMDC layers are nearly aligned. For $\theta^{(t)}=1^{\circ}$ one finds  $\lambda_{sm}^{(1)}=18$nm corresponding to  $E_M=0.13$eV, which is
comparable to the  ones found in hBN/Graphene/hBN. 
%While this energy is still much larger than the energy scale of the induced SOC, it may be relevant for transport studies....
Based on these results  one may expect an interesting interplay of induced SOC and supermoir\'e effects in  TMDC/graphene/TMDC for  $|\theta^{(b)}-\theta^{(t)}|\lesssim 1^{\circ}$.

%%%%%%%%%%%%%%%%%%%%%%%%%%%%%%%%%%%%%%%%%%%%%%%%%%%%%%%%%%%%%%%%%%%%%%%%%%%%%%%%%%%%%%%%%%%%%%%%%%%%%%%%%%%%%%%%%%%%%%%%%%%%%%%%%%%%%%%%%%%%%%%%%%%%%%%%%%%%%%%%%%%%%%%%%%%%%%%%%%%

\section{Magnetotransport calculations for graphene $npn$ junctions}
\label{sec:npn-junction}

\begin{figure}[ht]
  \includegraphics[scale=0.4]{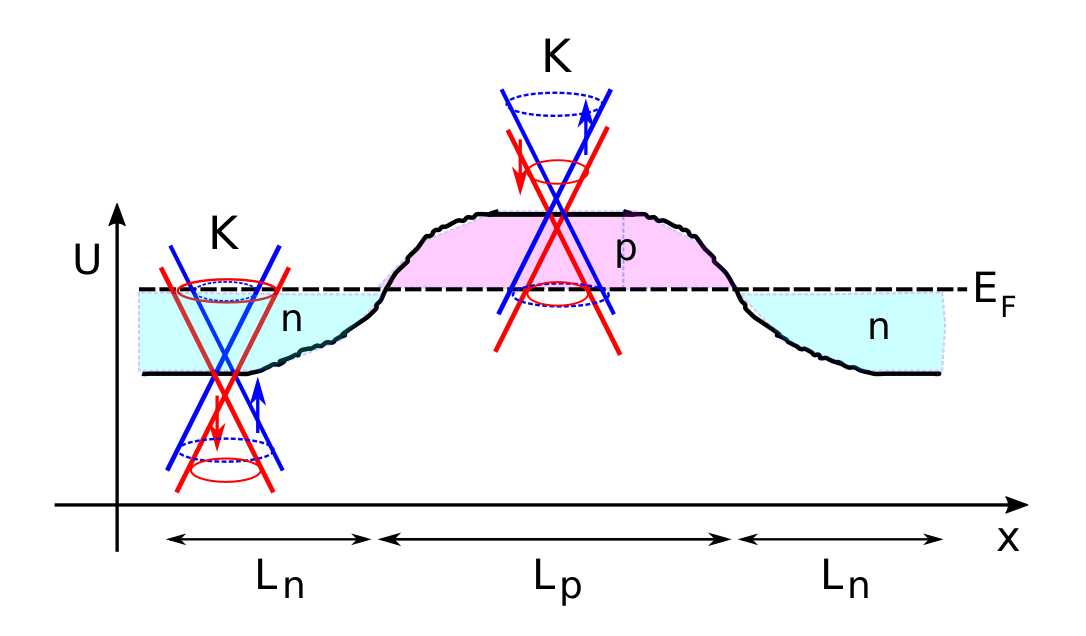}
  \caption{Schematics of a $npn$ junction in graphene. The Dirac point of graphene is shifted in energy due to the 
           doping potential $U(x)$ in the $n$ and $p$ doped regions. The length of the  $n$ ($p$) doped region is $L_n$( $L_p$).  $E_F$ denotes the Fermi energy. 
  \label{fig:Gr_npn_setup}}
\end{figure}

In this section we briefly explain the theoretical model that was used in the magnetotransport calculations through a $npn$ graphene junction with proximity induced SOC shown
in Fig.~4 of the main text.
We assume  that the  graphene flake is perfectly ballistic and due to external gates it is doped such that it hosts an $npn$ junction, see Fig.~\ref{fig:Gr_npn_setup}. 
We numerically calculate the transmission through the junction as a function of perpendicular magnetic field magnetic field $B_z$  and Fermi energy $E_F$ using the in-house code EQuUs\cite{equus-code}. 

We use the nearest-neighbor  tight-binding (TB) model of graphene: 
\begin{equation}
\hat{H}_{orb} = \sum_{i,s} U_i^{} \left(a_{i,s}^{\dagger} a_{i,s} + b_{i,s}^{\dagger} b_{i,s} \right)- 
\gamma_{}^{} \sum_{\langle ij\rangle} e^{\phi_{ij}} a_{i,s}^{\dagger} b_{j,s}+ h.c.
\end{equation}
where  $a_{i,s}^{\dagger}$ ($a_{i,s}$) and $b_{i,s}^{\dagger}$ ($b_{i,s}$) 
are creation (annihilation) operators for electrons of spin $s$ on the $A$ and $B$  sublattice site, respectively.  
$\gamma_{}=2.97\,$eV is the hopping amplitude between the nearest-neighbor atomic sites 
$\langle ij \rangle $ in the graphene lattice, $\phi_{ij}$ encodes the effect of perpendicular magnetic field through Peierls substitution, 
and  $U_i$  is the on-site energy on the atomic site $i$.
Following Ref.~\cite{CPR-graphene}, we used 
\begin{equation}
 U(x)=U_n+\frac{U_p-U_n}{2}\left(\tanh\left[\frac{x-L_n}{l_1}\right]-\tanh\left[\frac{x-L_n-L_p}{l_2}\right] \right).
\end{equation}
Here $L_n= 50$nm gives the length of the  $n$-doped regions and we used $U_n=-240\,$meV.  The  $p$-doped middle region  was $L_p=150\,$nm long and the doping was  $U_p=40\,$meV.
The parameters $l_1=l_2=25\,$nm  set the smoothness of the transition between the $n$ and $p$-doped regions.

In the TB formalism  the  Bychkov-Rashba SOC can be written as\cite{kane_quantum-spin-Hall_2005,frank_protected_2018}:
\begin{equation}
 H_{R} = i\,\frac{\lambda_R(\theta)}{3} e^{i \frac{s_z}{2} {\vartheta_{R}}} \sum\limits_{\left\langle i,j\right\rangle,s,s'}
 \left[{a}^{\dagger}_{i,s}\left(\boldsymbol{s}_{}\times
\mathbf{\widehat{d}}_{\left\langle i,j\right\rangle} \right)_z {b}_{j,s'} -  h.c. \right] e^{-i \frac{s_z}{2} {\vartheta_{R}}}\;.
\label{eq:H_RSO}
\end{equation}%
where $\boldsymbol{s}= (s_x,s_y,s_z)$ are the Pauli matrices representing the electron spin operator.
Moreover, $ \mathbf{\widehat{d}}_{\left\langle i,j\right\rangle} = \mathbf{d}_{\left\langle i,j\right\rangle}/d $
are unit vectors, where $\mathbf{d}_{\left\langle i,j\right\rangle}$
points from atom $j$ to its nearest neighbors $i$ and $d=|\mathbf{d}_{\left\langle i,j\right\rangle}|$.
The corresponding continuum SOC Hamiltonian, that can be obtained by Fourier transforming Eq(\ref{eq:H_RSO}) and expanding it 
at the $\pm K$ points of the Brillouin zone, reads 
$
H_{R}= \frac{\lambda_R(\theta)}{2}  e^{i \frac{s_z}{2} {\vartheta_{R}}} 
  \left(\tau_z\sigma_x s_y - \sigma_y s_x\right)  e^{-i \frac{s_z}{2} {\vartheta_{R}}}.
$  
As long as only a perpendicular magnetic field is applied, the phase $\vartheta_{R}$ will not affect the results, because 
the corresponding terms in Eq.~(\ref{eq:H_RSO}) can be removed by a unitary transformation.

The valley-Zeeman SOC can be written as\cite{frank_protected_2018}:
\begin{equation}
 H_{vZ} = \frac{i}{3\sqrt{3}}\,\sum\limits_{\left\langle\langle i,j\right\rangle\rangle s,s'} \left[
 \lambda_{vZ}^{(A)}\nu_{ij}[s_z]_{s,s'}{a}^{\dagger}_{i,s} {a}^{}_{j,s'}+\lambda_{vZ}^{(B)}\nu_{ij}[s_z]_{s,s'}{b}^{\dagger}_{i,s} {b}^{}_{j,s'}\right]\;.
 \label{eq:H_vZ}
\end{equation}%
It couples same spins and depends on  clockwise ($\nu_{ij}=1$) or counterclockwise ($\nu_{ij}=-1$) paths along a hexagonal ring from site $j$ to $i$. 
In the case of valley-Zeeman SOC  $\lambda_{vZ}^{(A)}=-\lambda_{vZ}^{(B)}$. The term Eq.(\ref{eq:H_vZ}) can also describe
the intrinsic SOC if the SOC strength $\lambda_{I}$ is the same on the two sublattices, i.e.,  $\lambda_{I}^{(A)}=\lambda_{I}^{(B)}$. 
The corresponding continuum SOC Hamiltonian reads $H_{vZ}=\tau_z s_z \lambda_{vZ}$.

We  used periodic boundary conditions perpendicular to the transport direction ($\hat{x}$ in Fig.\ref{fig:Gr_npn_setup}). 
%We can therefore assume that 
Thus, the transverse momentum $q_n$ is a good quantum number and the total transmission can be calculated as a sum over all $q_n$:
\begin{equation}
 T(E_F, B_z) = \sum\limits_{n} t(q_n, E_F, B_z) \;,
\end{equation}
where $t(q_n, E_F, B_z)$ is the transmission coefficient for mode $n$. 
When there are many transverse momenta the exact form of the boundary conditions is not important and therefore we 
used the infinite mass boundary condition\cite{graphene-boundary-cond} to obtain $q_n$:
%\begin{equation}
$
 q_n = \left(n+\frac{1}{2}\right)\frac{\pi}{W}\;,
%\end{equation}
$
where $n=0,1,2,\dots$ and $W$ is the width of the junction. %The conductance through the junction is then given by $G=\frac{e^2}{h}  T(E_F, B_z)$.

%%%%%%%%%%%%%%%%%%%%%%%%%%%%%%%%%%%%%%%%%%%%%%%%%%%%%%%%%%%%%%%%%%%%%%%%%%%%%%%%%%%%%%%%%%%%%%%%%%%%%%%%%%%%%%%%%%%%%%%%%%%%%%%%%%%%%%%%%%%%%%%%%%%%%%%%%%%%%%%%%%%%%%

\section{Electron scattering at a planar junction in a  graphene-WSe$_2$ system }
\label{sec:planar-junction}

We will discuss a further setup to investigate twist-angle dependent transport in a graphene-TMDC system. It  is motivated by 
recent electron optics experiment of Ref.\cite{barnard_absorptive_2017,bhandari_imaging_2018} where  collimated electron beams were created in monolayer graphene. 
By making use of such collimated electron beams one can test the twist-angle dependence of the induced SOC through spin-dependent scattering.

\begin{figure}
\begin{center}
	\includegraphics[scale=0.2]{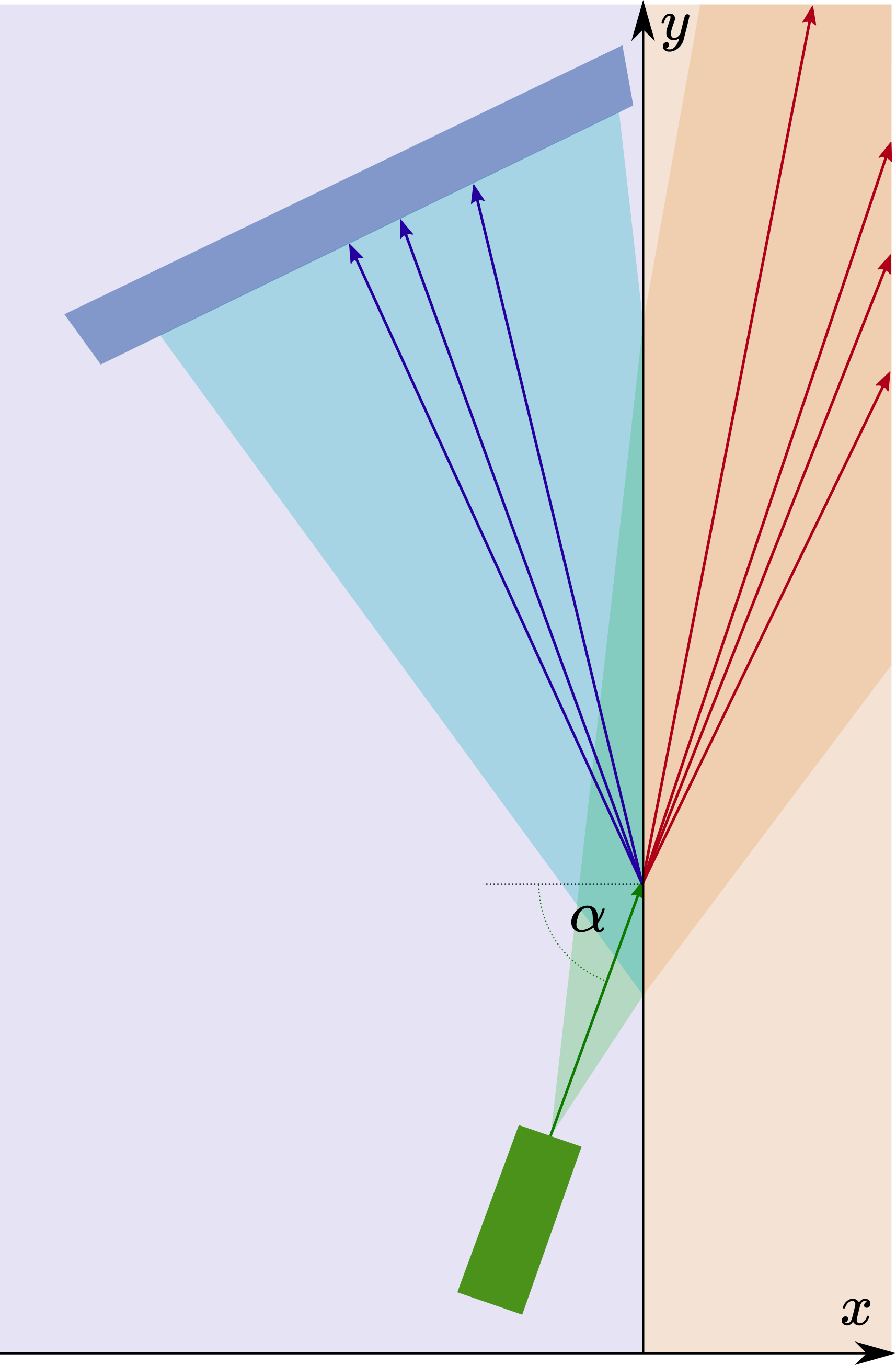}
	\caption{Ballistic electron scattering in MLG/WSe$_2$-WSe$_2$/MLG/WSe$_2$ junction. The top layer WSe$_2$ is present only on the right hand side (orange). 
	 The electron source produces a collimated beam of $20$\,meV electrons (green), which scatters into different modes at the junction. Transmitted electrons (red) get absorbed 
	 at the far side  of the system, while reflected electrons (blue) are absorbed by the drain electrode shown in blue. The reflection coefficient of the junction 
	  depends on the  rotation angle $\theta^{(t)}$ of the top layer on the right-hand side}
	\label{fig:setup}
	\end{center}
\end{figure}

To demonstrate this, we consider the TMDC/MLG/TMDC stack shown in Fig.~\ref{fig:setup}. It consists of a MLG/WSe$_2$ bilayer of a fixed twist angle $\theta^{(b)}$ (purple region). 
Part of this bilayer  is covered by a top WSe$_2$ layer (orange region) of twist angle $\theta^{(t)}$. Thus the induced SOC in graphene  changes at the  bilayer-trilayer boundary and in the trilayer region it is assumed to be tunable by  $\theta^{(t)}$. 
The left edge of the top layer defines a junction along the $y$ axis, where ballistic electrons scatter into various forward-propagating and back-propagating modes,
as indicated in Fig.~\ref{fig:setup}. We calculate below the reflection coefficient of this junction for a collimated beam of  electrons.

The general form of the Hamiltonian is $H^{gr}_{eff}=H_{orb}^{gr}+H_R^{gr}+H_{vZ}^{gr}$, see Sec.~\ref{sec:layer-rotation} for the definition of each term. 
On the left-hand side of the junction $H_R^{gr}$ is given in Eq.~(\ref{eq:H_R-transformed}), with $\theta=\theta^{(b)}$ and 
$\lambda_{vZ}=\lambda_{vZ}^{(b)}(\theta^{(b)})$  in $H_{vZ}^{gr}$.  On the right hand side, as shown in the main text,   the Rashba Hamiltonian is 
%\begin{equation}
 $H_{R}^{(tls)}
=\frac{\lambda^{(tls)}_R}{2} 
 e^{i \frac{s_z}{2} {\vartheta^{(tls)}}}
\left(\tau_z\sigma_x s_y - \sigma_y s_x\right)
e^{-i \frac{s_z}{2} {\vartheta^{(tls)}}},$
%\end{equation}
while $\lambda_{vZ}=\lambda_{vZ}^{(b)}(\theta^{(b)})+\lambda_{vZ}^{(t)}(\theta^{(t)})$ in $H_{vZ}^{gr}$. 
One can easily show  that the reflection coefficient is the same for electrons in both valleys,  for concreteness,  we perform the calculation for the $K$ valley.
We express the electronic states as plane waves multiplied by spinors in the basis 
$\left \{ |A\uparrow\rangle, |B\uparrow\rangle, |A\downarrow\rangle, |B\downarrow\rangle \right\}$ and we  assume that their energy is  $E=20$\,meV. 
Due to translational symmetry along $\hat{y}$ axis, the $\delta k_y$ component of the wavevector of the electronic states  is conserved. 
The SOC leads to the splitting of the bands, therefore  (at the $E_F$ given above) for each direction of propagation one  can find two modes for the incoming electrons 
with slightly different wave vectors and opposite pointing spins. We assume a net zero spin polarization with an equal mixture of the two incoming modes,  
and for each incoming mode, one can find two reflected and two refracted solutions. In total this gives four back-scattered and four forward-scattered modes  within one valley. 
Depending on the exact parameters and the angle of incidence, some of these modes can be decaying waves carrying no current. 
The particle current operator can be calculated as
$\mathcal{J}_x = \frac{i}{\hbar} [\mathcal{H}, x] %=\tau v_F s_0 \otimes \sigma_x [\nabla_x, x] 
= \tau v_F s_0 \otimes \sigma_x,
$
and in a similar fashion,
$
\mathcal{J}_y = v_F s_0 \otimes \sigma_y.
$
The ratio of the current carried by all the back-scattered waves and that of all the incoming modes give us the reflection coefficient: 
\begin{equation}
 R = - \sum_{i=1}^4 \langle\mathcal{J}_{x,i}^\text{(back)}\rangle \left/ \sum_{i=1}^2 \langle\mathcal{J}_{x,i}^\text{(in)}\rangle \right. .
\end{equation}

In an experimental scenario  one can never create a beam of electrons with an exact direction of propagation. Nevertheless, as shown in 
Refs.\cite{barnard_absorptive_2017, bhandari_imaging_2018} there are experimental techniques to collimate electrons in graphene from a source.  In Ref~\cite{barnard_absorptive_2017}
absorptive pinhole collimators were devised to collimate ballistic electrons  into a triangular shaped angular distribution with a half width at half maximum (HWHM) 
of just below $10^\circ$; while in Ref~\cite{bhandari_imaging_2018}  a cosine-shaped distribution was  achieved with a similar HWHM of $9^\circ$ using zigzag contacts to absorb stray electrons. Therefore we consider the effective reflection index $\overline{R}$ that is obtained by averaging $R$ with a certain normalized distribution $d$ centered 
around an angle of incidence $\alpha$:
\begin{equation}
 \overline{R}(\alpha) = \int d(\alpha'-\alpha) R(\alpha') \text{d}\alpha'.
\end{equation}

\begin{figure}
 \begin{center}
	\includegraphics[scale=0.25]{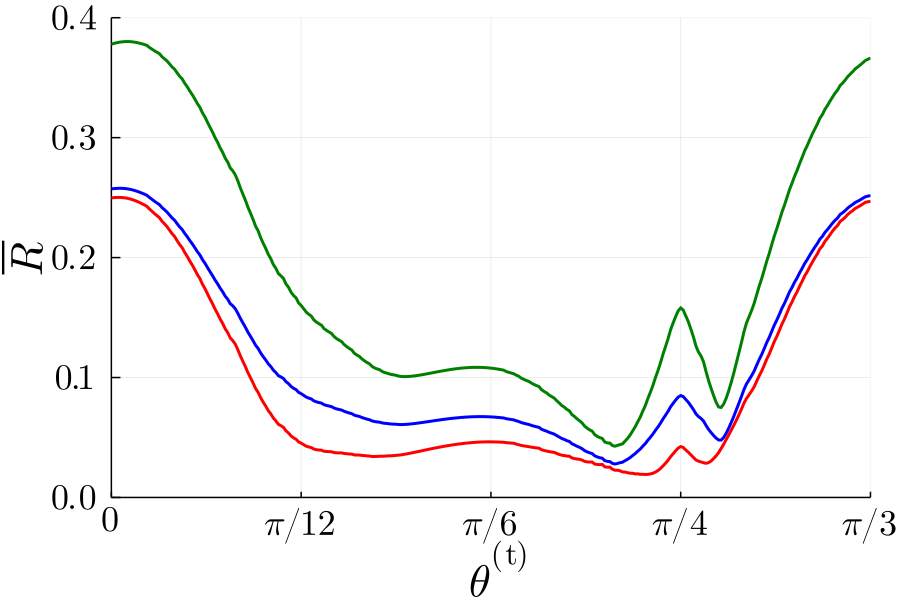}
	\vspace{0.3cm}
	\caption{The effective reflection $\overline{R}$ as a function of $\theta^{(t)}$.  Angular distribution of incoming electrons taken from Ref~\cite{barnard_absorptive_2017} (blue) 
	 and from Ref~\cite{bhandari_imaging_2018} (red) centered at $\alpha=70^\circ$. The green curve shows the result   for  $\alpha=75^\circ$
	 using the angular distribution from Ref~\cite{bhandari_imaging_2018}.
	}
	\label{fig:reflection}
	\end{center}
\end{figure}

In Fig~\ref{fig:reflection}, $\overline{R}$ is plotted as function of the rotation angle $\theta^{(t)}$ of the top TMDC layer for fixed twist angle  $\theta^{(b)}=\pi/4$  
of the bottom layer. As shown in Fig.2 of the main text, by changing $\theta^{(t)}$  all three SOC parameters $\lambda_R^{(tls)}$, $\vartheta^{(tls)}$ and $\lambda_{vZ}^{(tls)}$ 
discussed for the trilayer case are changing  and they all affect  $\overline{R}$. 
As one can see, the  main features in $\overline{R}$  do not seem to depend much on the specific angular distribution of the incoming plane waves or on the exact value of the central angle of incidence. 

Experimentally, in order to check the  predictions in Fig.~\ref{fig:reflection}, a  straightforward approach would be to prepare a few different samples with different $\theta^{(t)}$.
Alternatively, a setup similar to Ref.~\cite{ribeiro-palau_twistable_2018} could also be feasible, whereby the top layer is rotated \textit{in situ}.

\bibliographystyle{unsrt}
\bibliography{bibliography-suppl}